\documentclass[epsf,prb,twocolumn,showpacs,nofootinbib]{revtex4-1}

\usepackage[pdftex]{graphicx}
\usepackage{dcolumn}
\usepackage{bm}
\usepackage{epsfig}
\usepackage{latexsym}
\usepackage{amsmath}
\usepackage{amsfonts}
\usepackage{amssymb}
\usepackage{color}
\usepackage{array}
\usepackage{framed}

\begin{document}

\title{Crystal-to-Fracton Tensor Gauge Theory Dualities}
\author{Michael Pretko, Zhengzheng Zhai, and Leo Radzihovsky}
\affiliation{Department of Physics and Center for Theory of Quantum Matter\\
University of Colorado, Boulder, CO 80309}
\date{\today}

\begin{abstract}
We demonstrate several explicit duality mappings between elasticity of two-dimensional crystals and fracton tensor gauge theories, expanding on recent works by two of the present authors.  We begin by dualizing the quantum elasticity theory of an ordinary commensurate crystal, which maps directly onto a fracton tensor gauge theory, in a natural tensor analogue of the conventional particle-vortex duality transformation of a superfluid.  The transverse and longitudinal phonons of a crystal map onto the two gapless gauge modes of the tensor gauge theory, while the topological lattice defects map onto the gauge charges, with disclinations corresponding to isolated fractons and dislocations corresponding to dipoles of fractons.  We use the classical limit of this duality to make new predictions for the finite-temperature phase diagram of fracton models, and provide a simpler derivation of the Halperin-Nelson-Young theory of thermal melting of two-dimensional solids.  We extend this duality to incorporate bosonic statistics, which is necessary for a description of the quantum melting transitions.  We thereby derive a hybrid vector-tensor gauge theory which describes a supersolid phase, hosting both crystalline and superfluid orders.  The structure of this gauge theory puts constraints on the quantum phase diagram of bosons, and also leads to the concept of symmetry enriched fracton order.  We formulate the extension of these dualities to systems breaking time-reversal symmetry.  We also discuss the broader implications of these dualities, such as a possible connection between fracton phases and the study of interacting topological crystalline insulators.
\end{abstract}
\maketitle

\tableofcontents

\section{Introduction}

\subsection{Overview}

Stimulated by the ever-growing class of unusual quantum materials which do not conform to the conventional Landau paradigms of Fermi liquids and symmetry breaking, much effort has been directed at exploring models that exhibit quantum phases with exotic fractionalized quasiparticles.  Recently, a new class of quantum phases of matter has been discovered, featuring quasiparticles with unusual restrictions on their mobility.  The first, and most famous, example of this phenomenon is the ``fracton" excitation.  These exotic particles are characterized by strict immobility in isolation, but they can often move through interaction with other particles.  More generally, there exist particles which move freely only along certain subspaces while being immobile in the transverse directions, exhibiting subdimensional behavior.  Fractons and other subdimensional particles were first seen in the context of certain exactly solvable lattice models, such as stabilizer code spin models and Majorana systems.\cite{chamon,bravyi,haah,cast,yoshida,haah2,fracton1,fracton2}  It was later realized that these new particles have a natural theoretical description in the language of tensor gauge theories, which exhibit restricted mobility due to an unusual set of higher moment charge conservation laws, such as conservation of dipole moment.\cite{sub,genem,higgs1,higgs2}  Rapid recent progress in the field has established connections with numerous other areas of physics, such as localization\cite{abhinav,screening,circuit}, gravity\cite{mach}, holography\cite{holo,holo2}, quantum Hall systems\cite{theta,matter}, hole-doped antiferromagnets\cite{polaron}, and deconfined quantum criticality\cite{deconfined}, among many other theoretical developments.\cite{williamson,sagarlayer,hanlayer,foliation,parton,slagle,bowen,nonabel,balents,
field,albert,fusion,correlation,simple,entanglement,spectra,compactify,bernevig,
multipole,fractonice,fractoncs,twistfol,lego,gaugefrac1,gaugefrac2}  We refer the reader to Reference \onlinecite{review} for a review of fracton physics.

While the exotic properties of fractons have been the subject of intense study, concrete physical realizations have remained elusive until recently, when two of the present authors demonstrated explicitly that the fracton phenomenon is realized in an ordinary two-dimensional quantum crystal.\cite{prl}  More specifically, we provided a direct mapping between the quantum elasticity theory of a two-dimensional crystal and a tensor gauge theory featuring fracton excitations, in a direct tensor analogue of conventional particle-vortex duality.\cite{dasgupta,fisher}  Thus we explicitly demonstrate that fractons are directly realized in the crystal in the form of disclination defects.  The characteristic immobility of fractons is thereby demystified in terms of known constraints on the mobility of lattice defects.  Importantly, however, the duality from elasticity does not just give a generic tensor gauge theory, but one with an additional global $U(1)$ symmetry arising from atom number conservation.  This symmetry leads to the extra feature of subdimensional dipoles (dislocations), as is known from elasticity theory at zero temperature.  In contrast, a generic tensor gauge theory does not exhibit this feature without endowing it with extra structure.

In the present work, we derive and analyze this mapping in more detail, putting the duality on firmer ground.  We show that this duality allows for a productive exchange of ideas between two heretofore disconnected fields.  For example, the well-studied phase diagram of elasticity theory allows us to map out new finite-temperature phases of the corresponding fracton theory, such as analogues of the hexatic and isotropic liquid phases of elasticity theory.  In turn, fracton tensor gauge theory provides a convenient language for encoding the restricted mobility of lattice defects and allows for a simpler description of two-dimensional thermal melting transitions.  We note that mathematically similar gauge duals of elasticity theory have been studied in the literature, without identification of fracton order.\cite{kleinert1,kleinert2,zaanen}  Our work also provides significant technical simplifications over previous duality formulations.

In addition to the properties of crystals, we can also use our duality approach to describe quantum melting transitions.  However, in this case, the statistics of the underlying atoms of the crystal become important.  For simplicity, we focus primarily on the case where the atoms of the crystal are bosonic, such that a quantum liquid phase will naturally have superfluid order.  (We also comment on possible extensions to the case of fermionic atoms.)  To this end, the tensor gauge theory description of crystals must be combined with conventional particle-vortex duality, which is capable of describing superfluidity.  The end result is a hybrid vector-tensor gauge theory which describes a nontrivial interplay between crystalline and superfluid order, thereby providing a natural dual description of a supersolid, as first described in Reference \onlinecite{prl2}.  Here we provide a more complete derivation of this duality and explore its various consequences.  By condensing various topological defects, a supersolid can be driven into superhexatic, superfluid, or commensurate solid phases.  Importantly, however, the structure of the gauge dual rules out the possibility of zero-temperature hexatic or liquid phases without superfluid order, consistent with conventional wisdom.  This more complete version of the duality teaches us important lessons about fracton physics, such as the role of symmetry enrichment in restricting the mobility of particles.  For example, the glide constraint of a commensurate crystal, which is relaxed in the supersolid phase, corresponds to the one-dimensional motion of dipoles in the presence of a global $U(1)$ symmetry.

We end with a discussion of various connections that these dualities draw between the study of fracton phases and other areas of condensed matter theory.  For example, the defects of crystalline order carry quantum numbers related to the superfluid order parameter, and vice versa, in close relation to the physics of deconfined quantum criticality.  We also show how this duality provides a possible connection between fracton physics and the classification of interacting topological crystalline insulators (TCIs).

\subsection{Summary of Results}

The primary result of this paper is a set of dualities connecting the physics of fractons to the elastic theory of two-dimensional crystals.

\begin{figure}[t!]
 \centering
 \includegraphics[scale=0.47]{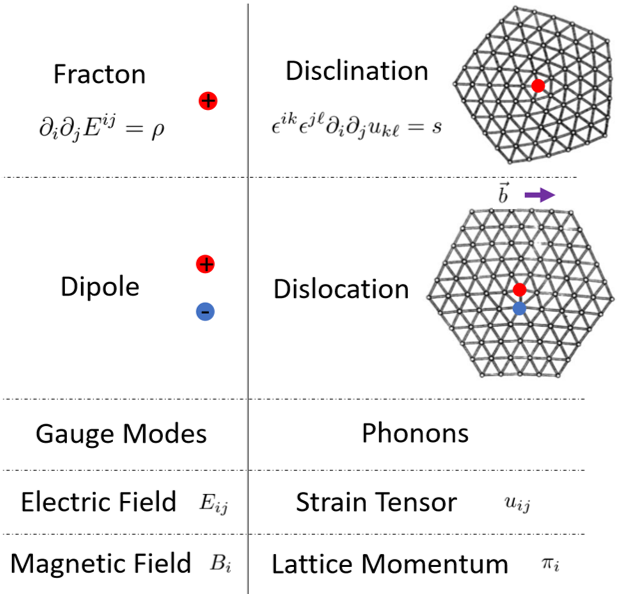}
 \caption{The excitations and operators of the scalar charge theory are in one-to-one correspondence with those of elasticity theory.  (Pictures of lattice defects adapted from Reference \onlinecite{seung}.)}
 \label{fig:dictionary}
 \end{figure}

The first duality we demonstrate starts from the standard elastic description of an ordinary commensurate crystal in terms of a displacement field $u^i(\vec{x})$, with action given by:
\begin{equation}
S = \int d^2xdt\frac{1}{2}\bigg((\partial_tu^i)^2 - C^{ijk\ell}u_{ij}u_{k\ell}\bigg).
\end{equation}
(All indices refer to spatial coordinates, and repeated indices are summed over.  Raising and lowering is done via a flat metric, $\delta^{ij}$.  The use of upper and lower indices is merely a bookkeeping device.)  This action is then mapped onto that of a tensor gauge theory coupled to fracton excitations, featuring a noncompact rank-two symmetric tensor gauge field $A_{ij}(\vec{x})$, along with a scalar potential $\phi(\vec{x})$:
\begin{equation}
S = \int d^2xdt\bigg(\frac{1}{2}\tilde{C}^{-1}_{ijk\ell}E^{ij}_\sigma E^{k\ell}_\sigma - \frac{1}{2}B^iB_i - \rho\phi - J^{ij}A_{ij}\bigg).
\end{equation}
where $E^{ij}_\sigma = -\partial_t A^{ij} - \partial^i\partial^j\phi$ and $B^i = \epsilon_{jk}\partial^jA^{ki}$.  Both of these actions feature two gapless modes with linear dispersion ($\omega\sim k$): phonons of the elastic theory and gauge modes of the gauge theory.  Additionally, each side of the duality hosts a set of topologically stable excitations: lattice defects of the crystal and charges of the gauge theory.  We show that these topological excitations of the two theories can be directly mapped onto each other.  In particular, disclinations (orientational lattice defects) correspond to isolated fracton charges, while dislocations (translational lattice defects) correspond to stable dipoles of fractons.  The derivation of this duality proceeds in close analogy with conventional particle-vortex duality, which maps the low-energy theory of a superfluid onto a conventional $U(1)$ gauge theory (as reviewed in Appendix A).  The full dictionary of fracton-elasticity duality is summarized in Figure \ref{fig:dictionary}.  We show how to derive this duality starting from either side: using the gauge theory to derive elasticity theory or vice versa.

As an important check on the validity of this duality, we explicitly demonstrate the equivalence of mobility restrictions on both the elastic and tensor gauge theory sides, as manifested in conservation laws and continuity equations.  For example, any motion of a disclination involves the absorption or creation of dislocation defects\cite{emit1,emit2,emit3,emit4}, just as motion of a fracton in the gauge theory involves absorption or emission of dipoles.  An isolated disclination will be strictly immobile, making it a true fracton excitation.  Similarly, dislocations can only move easily via gliding (motion in the direction of the Burgers vector).  In contrast, dislocation climb (motion perpendicular to the Burgers vector) involves the absorption/emission of another class of lattice defects: vacancies and interstitials.  In the absence of such auxiliary defects, an isolated dislocation moves only along its Burgers vector.  We show that the dipoles of the fracton gauge theory have the same quasi-one-dimensional behavior after taking into account symmetry quantum numbers related to the underlying atoms.  The gauge dual is thereby seen to be a symmetry enriched fracton phase, in which extra mobility restrictions are enforced by the presence of a global $U(1)$ symmetry, associated with atom conservation.

This fracton-elasticity duality not only provides numerous insights into the emerging field of fractons by drawing on established results of elasticity theory, but also allows fracton physics to shed new light on old problems of elasticity.  The restricted motion of fractons is put on more familiar grounds in terms of the known constraints on motion of lattice defects.  In turn, the conservation laws of higher rank tensor gauge theories provide a convenient language for systematically encoding such mobility restrictions.  Furthermore, using the well-studied phase diagram of classical two-dimensional elasticity theory, we can map out several new classical (finite-temperature) phases of the corresponding fracton model, including gauge theory equivalents of the hexatic and isotropic liquid phases.  We can then use the gauge theory formulation of fractons to study the melting transitions of two-dimensional crystals.  Specifically, we use the classical limit of our duality to provide a simpler analysis of the two-stage thermal melting transitions of two-dimensional crystals, first investigated by Halperin and Nelson and by Young.\cite{halperin,nelson,young}

While the duality transformation described above is sufficient for understanding classical melting, it fails to accurately capture the physics of quantum melting transitions.  By not accounting for the quantum statistics of the underlying atoms, it fails to capture important physics, such as the fact that a liquid of bosonic atoms at zero temperature should form a superfluid, instead of a truly featureless state.\cite{lsm,hastings,oshikawa}  In order to rectify this deficiency, we must formulate a more complete gauge dual which combines the properties of both fracton-elasticity and particle-vortex duality, thereby allowing for simultaneous description of crystalline and superfluid orders.

To this end, we start from a low-energy field theory description of a supersolid, featuring both types of order, with an action given by:
\begin{eqnarray}
S = \int_{x,t}\bigg[\frac{1}{2}\rho(\partial_t u_i)^2 - \frac{1}{2} C^{ijk\ell} u_{ij} u_{k\ell}
+\frac{1}{2}\chi (\partial_t\varphi)^2 \nonumber \\
- \frac{1}{2} K (\partial_i\varphi)^2 - g_1\partial_t u^i\partial_i\varphi
+ g_2\partial_t\varphi\partial_i u^i\bigg],
\end{eqnarray}
where $u_i$ is again the lattice displacement and $\varphi$ is the phase of the superfluid condensate.  The various terms and parameters of this action will be discussed in detail later.  Note that the last two terms represent nontrivial coupling between the superfluid and crystalline sectors, which has important consequences in the dual description.  After performing an appropriate duality transformation, the supersolid action maps onto the following hybrid vector-tensor gauge theory:
\begin{align}
S =&\int_{x,t}\bigg[\frac{1}{2}\hat C_{ijk\ell} E^{ij}_\sigma E^{k\ell}_\sigma -
\frac{1}{2}\bar\rho^{-1} B^2+\frac{1}{2}\bar K^{-1}e^2 - \frac{1}{2}\bar\chi^{-1} b^2\nonumber\\
&- \bar g {\bf B}\cdot  {\bf e} - \underline{g} E^{ii}_\sigma b - J_s^{ij} A_{ij} - s A_0 - {\bf j}_v\cdot  {\bf a} - n_v a_0\bigg].
\end{align}
where $E^{ij}_\sigma$ and $B^i$ are defined as previously, while $e^i = -\partial_ta^i - \partial^ia_0$ and $b = \epsilon^{ij}\partial_ia_j$.  Note that, in addition to Maxwell-type ``$E^2+B^2$" terms for the vector and tensor gauge fields, the action also features cross terms coupling the electric field of one sector to the magnetic field of the other sector.  These pieces of the action can be interpreted as generalized axion contributions, analogous to the $E\cdot B$ term featured in axion electrodynamics.\cite{witten,wilczek}  As in conventional axion physics, we expect that this coupling will result in a form of charge attachment.  Indeed, we show that the effect of these cross terms is to attach quantum numbers of one sector to the gauge charges of the other.  Specifically, the topological lattice defects of crystalline order carry boson number of superfluid sector, while vortices of the superfluid order carry angular momentum of the crystalline order.

This relationship between the two orders has important consequences for the quantum ($i.e.$ zero-temperature) phase diagram of bosons.  For example, when a commensurate crystal undergoes a quantum melting transition via condensation of topological lattice defects, the underlying bosons necessarily condense as well, leading to superfluid order.  Similarly, condensation of vortices in the superfluid phase will automatically cause the system to form crystalline order.  In this way, we find that it is impossible for a system of bosons in the continuum to have a trivial gapped state at zero temperature, which is consistent with both conventional wisdom and the Lieb-Schultz-Mattis theorem.\cite{lsm,hastings,oshikawa}  We can also further conclude that even the partially melted hexatic phase, obtained from a solid via proliferation of dislocation defects, must necessarily feature superfluid order, ensuring that a non-superfluid hexatic phase does not exist at zero-temperature.  We use this insight to establish the full quantum phase diagram of intertwined superfluid and crystalline orders.

Finally, we discuss possible connections between the duality established here and other topics in condensed matter physics.  For example, the relationship between quantum numbers of defects of the two sectors draws an immediate connection with the physics of deconfined quantum criticality, in which continuous quantum phase transitions are allowed between phases with different order parameters via similar interplay of topological defects.\cite{dqcp,defined}  As such, the gauge dual of the supersolid discussed here draws a connection with the theory of deconfined quantum critical points.  As another application, we note that recent connections have been drawn between the theory of lattice defects and topological crystalline insulators (TCIs).\cite{else}  Our duality suggests that the full characterization of fracton phases (which does not currently exist but is a subject of active research) will be an important tool for the classification of interacting TCIs.  More generally, this duality will allow for a productive future exchange of ideas between the new field of fractons and established literature in the field of elasticity.

\section{Background}

\subsection{Two-Dimensional Elasticity Theory}

In this work, we will primarily focus on the elastic theory of a two-dimensional quantum crystal, in which the underlying atoms have arranged themselves into a lattice with translational and orientational order.  Each atom can oscillate only a small distance $u^i$ away from its equilibrium position, which serves as the fundamental dynamical variable of elasticity theory.  Note that the system remains invariant under a global shift of $u^i$, indicating that the low-energy theory must only involve derivatives of $u^i$.  To linear order, the most general low-energy action that can be written down is\cite{chaikin,landau,kardar}:
\begin{equation}
S = \int d^2xdt\frac{1}{2}\bigg((\partial_tu^i)^2 - C^{ijk\ell}u_{ij}u_{k\ell}\bigg),
\label{elastact}
\end{equation}
where $u_{ij}$ is the symmetrized strain tensor:
\begin{equation}
u_{ij} = \frac{1}{2}(\partial_iu_j + \partial_ju_i).
\end{equation}
In writing this action, we have implicitly assumed time-reversal symmetry, as we will do throughout most of this manuscript, unless indicated otherwise.  Note that the antisymmetric  component of the strain tensor, or equivalently the bond angle, $\theta_b = \frac{1}{2}\epsilon^{ij}\partial_iu_j$, cannot appear explicitly in the action, as a consequence of the underlying rotational symmetry (which is spontaneously broken in the crystal).\cite{chaikin}  Rather, only derivatives of the bond angle can appear in the action, which are irrelevant contributions with subdominant effect on the low-energy dispersion.  In Section \ref{sec:alt}, we will describe an alternative formulation of elasticity theory which treats the bond angle more explicitly.  For now, we focus on the action in terms of the symmetric strain tensor, which leads to a linear gapless dispersion ($i.e.$ $\omega\sim k$) for the two components of $u^i$, corresponding to transverse and longitudinal phonons.

\begin{figure}[t!]
 \centering
 \includegraphics[scale=0.25]{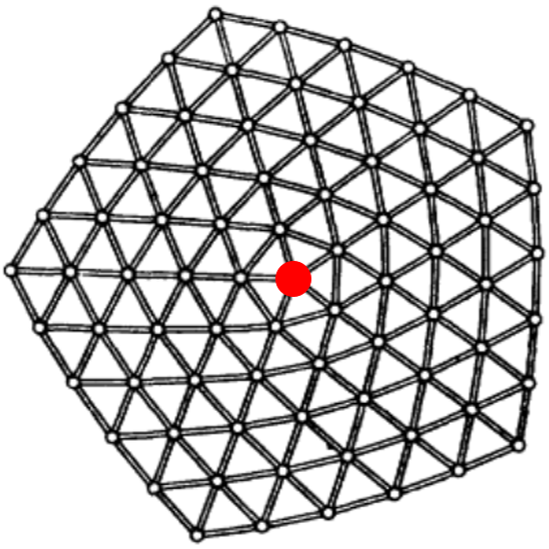}
 \caption{Disclinations are orientational defects of the crystal, as depicted above on the triangular lattice.  Notice that the central site touches only five other sites, indicating a missing bond angle of $\pi/3$.  (Figure adapted from Reference \onlinecite{seung}.)}
 \label{fig:disc}
 \end{figure}

While the bond angle $\theta_b$ makes no appearance in the action, it is still useful for defining disclinations, the fundamental topological defects of a crystal, which represent defects of the orientational order of the system, as depicted in Figure \ref{fig:disc}.  Going around a path enclosing a fundamental disclination of a two-dimensional crystal with $C_n$ symmetry, the bond angle $\theta_b$ will change by $2\pi/n$.  In equations, this corresponds to:
\begin{equation}
\oint d\ell^i\,\partial_i\theta_b = \frac{2\pi}{n}s,
\end{equation}
where $d\ell^i$ is tangent to the curve, and $s$ is an integer representing the total number of enclosed disclinations.  It is also useful to rewrite the disclinations in terms of the symmetric strain $u_{ij}$ as follows:
\begin{align}
\begin{split}
\frac{2\pi}{n}s &= \oint d\ell^i\,\partial_i\theta_b = \frac{1}{2}\oint d\ell^i\,\epsilon^{kj}\partial_k(\partial_iu_j + \partial_ju_i - \partial_ju_i) \\
 &= \oint dn_\ell\epsilon^{i\ell}\epsilon^{kj}(\partial_ku_{ij} - \frac{1}{2}\partial_k\partial_ju_i)\\
 &= -\int d^2x\,\epsilon^{i\ell}\epsilon^{jk}\partial_\ell\partial_ku_{ij} - \frac{1}{2}\int d^2x\,\epsilon^{i\ell}\partial_\ell(\epsilon^{kj}\partial_k\partial_j u_i),
\end{split}
\end{align}
where $dn^i = \epsilon^{ji}d\ell_j$ is normal to the curve, and we have integrated by parts via Stokes theorem and have freely commuted derivatives on the boundary, away from any singularities.  We now define a disclination density $\rho_s$ via:
\begin{equation}
\rho_s = \epsilon^{i\ell}\epsilon^{jk}\partial_\ell\partial_ku_{ij},
\end{equation}
which allows us to write the total disclination number as:
\begin{equation}
-\frac{2\pi}{n}s = \int d^2x\,\bigg(\rho_s - \epsilon_{i\ell}\partial^i(\frac{1}{2}\epsilon^{kj}\partial_k\partial_ju_i)\bigg).
\end{equation}
The first term above corresponds to the bare disclination density of the system.  In order to make sense of the second term, we must take into account a second type of topological defect found in two-dimensional crystals.

\begin{figure}[t!]
 \centering
 \includegraphics[scale=0.22]{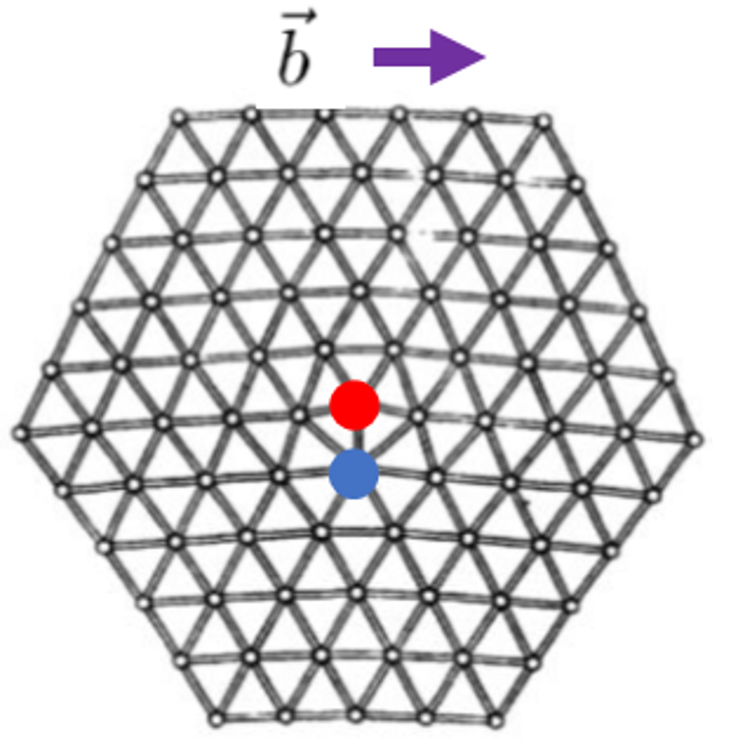}
 \caption{Dislocations correspond to bound states of two equal and opposite disclinations, representing translational defects of the crystal.  Note that the Burgers vector $\vec{b}$ is perpendicular to the vector between the two disclinations.  (Figure adapted from Reference \onlinecite{seung}.)}
 \label{fig:disl}
 \end{figure}

In addition to the fundamental disclinations, a two-dimensional crystal supports topological defects corresponding to bound states of two equal and opposite disclinations, as shown in Figure \ref{fig:disl}.  This bound state, which is a dipole of disclinations, corresponds to a defect of the translational order of the system.  Upon going around a curve enclosing the defect, $u^i$ changes by a constant $b_j$, known as the Burgers vector, which is perpendicular to the line between the two constituent disclinations.  Note that the Burgers vector is constrained to be a lattice vector.  In equations, we can write:
\begin{equation}
\oint d\ell^i\,\partial_iu_j = b_j.
\end{equation}
It is also possible to rewrite this equation purely in terms of the symmetric strain, $u_{ij}$, taking advantage of the fact that dislocations correspond to dipolar bound states of equal and opposite disclinations.\cite{chaikin,landau,seung,sarang}  Assuming that we are considering a region with zero net disclinations, we can integrate by parts twice (and relabel several indices) to write:
\begin{align}
b_n &= \epsilon_{mn}\int d^2x\,(\rho x^m) = \int d^2x\,x^m\epsilon_{mn}\epsilon^{i\ell}\epsilon^{jk}\partial_\ell\partial_ku_{ij}\nonumber\\
&=\oint d\ell^i x^m\epsilon_{mn}\epsilon^{jk}\partial_k\partial_iu_j + \int d^2x\,\rho_{b,n}\nonumber\\
&=\oint d\ell^i\epsilon_{in}\epsilon^{jk}\partial_ku_j + \oint d\ell^i\partial_iu_n = \oint d\ell^i\partial_iu_n,
\end{align}
where we have defined the dislocation density $\rho_b^n = \epsilon^{ik}\partial_k\partial_iu^n$, and in the last line we have assumed $\theta_b = \frac{1}{2}\epsilon^{jk}\partial_ju_k$ is single-valued, which amounts to assuming zero net disclination charge in the region.  In this sense, we can directly identify dislocations as dipoles of disclinations.  Notice that a dislocation is still a stable defect, exhibiting a topological winding, despite being ``neutral" in terms of the fundamental disclinations.  This fact will find a natural interpretation in the dual gauge theory.  Using the variables we have now defined, we can write the total disclination charge as:
\begin{equation}
-\frac{2\pi}{n}s = \int d^2x\,(\rho_s - \epsilon_{i\ell}\partial^i\rho_b^\ell)
\end{equation}
In this language, we see that $\rho_s$ represents bare disclinations, while the second term above represents the contribution to the total disclination density arising from the dislocation density $\rho_b$.

\subsection{Fracton Tensor Gauge Theory}

We now describe the appropriate fracton tensor gauge theory, known as the ``scalar charge theory" in the fracton literature, which we will see has strikingly similar physics to two-dimensional elasticity.  We here review the essential features of the theory, referring the reader to previous literature for a more detailed treatment.\cite{sub,genem}  The dynamical gauge variable of this theory is a rank-two symmetric tensor gauge field, $A_{ij}$, along with its canonical conjugate variable, which we denote as $E_{ij}$, playing the role of a generalized electric field tensor.  The gauge theory can be fully defined by specifying the gauge transformation, then writing the most general gauge-invariant low-energy action.  As discussed in previous references\cite{sub,genem}, an equivalent way to define the theory is to specify the generalized Gauss's law, which in turn determines the gauge symmetry.  For the scalar charge theory, the Gauss's law takes the form:
\begin{equation}
\partial_i\partial_j E^{ij} = \rho,
\end{equation}
for scalar charge density $\rho$, where repeated indices are summed over.  Since $E^{ij}$ is conjugate to $A^{ij}$, this Gauss's law immediately dictates that the low-energy sector is invariant under the following gauge transformation\cite{sub}:
\begin{equation}
A_{ij}\rightarrow A_{ij}+\partial_i\partial_j\alpha,
\end{equation}
for scalar $\alpha$ which is an arbitrary function of space.

As with more conventional gauge theories, this gauge structure leads to conservation of charge.  One simple way to see this is to consider the total charge $q$ contained in some region $V$ with boundary $\partial V$:
\begin{equation}
q = \int_V d^2x\,\rho = \int_V d^2x\,\partial_i\partial_jE^{ij} = \int_{\partial V} dn_i\,\partial_j E^{ij},
\end{equation}
where $dn_i$ is the normal vector on the boundary.  Just as in conventional electromagnetism, the total charge is encoded as a flux through the boundary.  As such, no local operator in the interior of $V$, far away from the boundary, can cause a change of $q$.  The charge of the system only changes when charges flow in or out through the boundary.

The more unusual aspect of this theory is that it also exhibits an additional conservation law, namely conservation of dipole moment.  Consider the total dipole moment $P^i$ contained in the region $V$:
\begin{align}
\begin{split}
P^i &= \int_V d^2x\,(\rho x^i) = \int_V d^2x\,x^i\partial_j\partial_kE^{jk} \\
&=\int_{\partial V} dn_j\,(x^i\partial_kE^{jk} - E^{ij}).
\end{split}
\end{align}
(Recall that the dipole moment is only independent of the choice of origin if the system is charge neutral.  Otherwise, the dipole moment can change by an overall constant depending on the origin choice.  In either case, all physical observables are independent of this arbitrariness of dipole definition.)  Unlike in conventional electromagnetism, we see that the dipole moment of this theory can be written as a flux encoded on the boundary, just like charge.  As such, no local operator in the interior of $V$ can cause a change of the total dipole moment.  Dipole moment only changes when charges pass through the boundary.  In other words, dipole moment is locally conserved.  This extra conservation law has dramatic consequences for the particles of the theory.  In particular, an isolated charge is strictly locked in place, since motion of a single charge would change the dipole moment of the system.  Only neutral bound states, such as dipoles, can move around the system.  These facts indicate that the fundamental charges of this theory meet the definition of fracton excitations.  Furthermore, the dipolar conservation law implies that the dipoles of the theory are topologically stable excitations, despite being charge-neutral.

Generically, the dipoles in a gauge theory of this form are completely mobile, unlike the disclination dipoles ($i.e.$ dislocations) in the context of elasticity theory, which are one-dimensional.  However, there is a simple way that subdimensionality can be incorporated into this theory via imposing a global symmetry.  Consider the trace of the total quadrupole moment tensor contained in a region $V$:
\begin{align}
Q^i_{\,\,i} &= \int_V d^2x\,\rho x^2 = \int_V d^2x\,x^2\partial_i\partial_jE^{ij} \nonumber\\
&= \int_{\partial V} dn_j(x^2\partial_iE^{ij} - 2x_iE^{ij}) + \int_V d^2x\,E^i_{\,\,i}.
\end{align}
Note that, up to boundary terms, this component of the quadrupole moment is equivalent to the integrated trace of the electric tensor, which is a more conventional global quantity ($i.e.$ without factors of $x$ in the integrand).  If we happened to have an ordinary global symmetry which required the integrated trace to vanish, or be a boundary term, then this quadrupole moment would automatically be conserved as well.  In a certain sense, this amounts to a higher moment conservation law being ``bootstrapped" to a conventional one, similar to the analysis of Reference \onlinecite{cheng}.  While such a global conservation law is not generically present in fracton theories, we will see that the gauge dual of elasticity theory has this type of conservation law, thereby making it a symmetry enriched fracton phase with reduced mobility.

In addition to gauge charges of restricted mobility, which we generically take to be gapped, this theory will also have gapless gauge modes, analogous to the gapless photon of conventional Maxwell theory.  To describe these modes, we write down the most general gauge-invariant Hamiltonian for the charge-free sector, which we can write in a form analogous to Maxwell theory:
\begin{equation}
H = \int d^2x\,\bigg(\frac{1}{2}\tilde{C}^{ijk\ell}E_{ij}E_{k\ell} + \frac{1}{2}B^iB_i\bigg),
\label{ham}
\end{equation}
where $\tilde{C}^{ijk\ell}$ is some matrix of coefficients.  (The precise number of independent coefficients will be dictated by the symmetries of the system.)  The magnetic field is a gauge-invariant operator given by $B^i = \epsilon_{jk}\partial^j A^{ki}$, describing the two physical components of $A_{ij}$.  As such, the equations of motion from this Hamiltonian yield two gapless gauge modes with linear dispersion, $\omega\sim k$.  By performing a canonical transformation, we can also rewrite this theory in the Lagrangian formalism, yielding the following action:
\begin{equation}
S = \int d^2xdt\bigg(\frac{1}{2}\tilde{C}_{ijk\ell}^{-1}E^{ij}_\sigma E^{k\ell}_\sigma - \frac{1}{2}B^iB_i\bigg),
\end{equation}
where $\tilde{C}_{ijk\ell}^{-1}$ is the matrix inverse of $\tilde{C}_{ijk\ell}$, such that $\tilde{C}_{ijk\ell}^{-1}\tilde{C}^{k\ell mn} = \delta_{ij}\delta^{mn}$.  We have also defined a new electric field quantity as:
\begin{equation}
E^{ij}_\sigma = -\partial_tA^{ij} - \partial^i\partial^j\phi,
\end{equation}
in which $\phi$ plays the role of a scalar potential function.  (The $\sigma$ notation will be explained in a subsequent section.)  Note that this new variable differs from the previous definition of electric field by a tensor factor:
\begin{equation}
E_{ij} = -\frac{\partial\mathcal{L}}{\partial \dot{A}^{ij}} = \tilde{C}^{-1}_{ijk\ell}E^{k\ell}_\sigma.
\end{equation}
The field $E^{ij}_\sigma$ is the analog of the electric displacement vector $\vec{D}$ in conventional electromagnetism.\cite{haldane}  The Lagrangian formalism of the theory is invariant under time-dependent gauge transformations, of the form:
\begin{equation}
A_{ij}\rightarrow A_{ij} + \partial_i\partial_j\alpha,\,\,\,\,\,\,\,\,\,\,\,\,\,\,\phi\rightarrow\phi +\partial_t\alpha,
\end{equation}
where $\alpha$ is now an arbitrary function of space and time.  Note that, after writing the physical fields in terms of gauge potentials, we can easily see that the following equation holds identically:
\begin{equation}
\partial_t B^i + \epsilon_{jk}\partial^j E_\sigma^{ki} = 0,
\end{equation}
which serves as the generalized Faraday's equation of the theory.

In the presence of fracton charges coupled to the gauge field, we must also add source terms to the action, resulting in:
\begin{equation}
S = \int d^2xdt\bigg(\frac{1}{2}\tilde{C}^{-1}_{ijk\ell}E^{ij}_\sigma E^{k\ell}_\sigma - \frac{1}{2}B^iB_i - \rho\phi - J^{ij}A_{ij}\bigg),
\end{equation}
where $\rho$ and $J_{ij}$ are the fracton charge density and fracton tensor current, which obey the following relationship, enforced by gauge invariance\cite{genem}:
\begin{equation}
\partial_t\rho + \partial_i\partial_j J^{ij} = 0,
\end{equation}
representing a generalized continuity equation.

\section{Fracton Gauge Dual of Commensurate Quantum Crystals}
\label{fgd}

\subsection{Derivation of the Duality}

We have now encountered two theories with essentially identical excitation spectra.  Both two-dimensional elasticity theory and the fracton tensor gauge theory exhibit two gapless gauge modes with linear dispersion, topological charge excitations, and stable dipoles.  We now demonstrate an explicit mapping between the two theories, starting from elasticity theory and deriving the fracton gauge theory.  In Appendix B, we will execute the duality in reverse, starting with the gauge theory to derive the elasticity theory.

The essential input that we need from two-dimensional elasticity theory is the action:
\begin{equation}
S = \int d^2xdt\,\frac{1}{2}\left[(\partial_t u^i)^2 -
C^{ijk\ell}u_{ij}u_{k\ell}\right],
\label{orig}
\end{equation}
given in terms of the symmetrized strain tensor, $u_{ij} = \frac{1}{2}(\partial_i u_j + \partial_j u_i)$, along with the source equation dictating how the strain responds to the presence of disclinations:
\begin{equation}
  \epsilon^{ik}\epsilon^{j\ell}\partial_i\partial_ju_{k\ell} = \rho.
\label{disc}
\end{equation}
Dislocations are also implicitly accounted for in this equation,
since a dislocation can be regarded as a bound state of two
disclinations,\cite{chaikin,landau,seung,sarang} as we will see explicitly.  In order to obtain the gauge dual, it is useful to first separate the displacement field into its singular and smooth phonon pieces, in terms of which we write $u_{ij} = u_{ij}^{(s)} + \frac{1}{2}(\partial_i\tilde{u}_j
+ \partial_j\tilde{u}_i)$, where $\tilde{u}_i$ is a smooth
single-valued function, obeying $\epsilon^{ik}\epsilon^{j\ell}\partial_i\partial_j\tilde{u}_{k\ell} = 0$.  The singular strain component
$u_{ij}^{(s)}$ represents the contribution from disclinations, $\epsilon^{ik}\epsilon^{j\ell}\partial_i\partial_ju_{k\ell}^{(s)} = \rho_s$.

We now introduce two Hubbard-Stratonovich fields, the lattice momentum $\pi_i$ and the stress tensor $\sigma_{ij}$.  In terms of these variables, we rewrite the action as:
\begin{align}
\begin{split}
S = \int d^2x dt\,&\bigg[\frac{1}{2}C^{-1}_{ijk\ell}\sigma^{ij}\sigma^{k\ell} - \frac{1}{2}\pi^i\pi_i \\
& - \sigma^{ij}(\partial_i \tilde{u}_j + u_{ij}^{(s)}) + \pi^i\partial_t (\tilde{u}_i + u_i^{(s)}) \bigg].\label{action2} 
\end{split}
\end{align}
This form for the action explicitly recovers Eq. \ref{orig} upon integrating out the fields $\pi_i$ and $\sigma_{ij}$.  Notice that the action is now linear in the smooth displacement field $\tilde{u}_i$, which can be integrated out to enforce the constraint:
\begin{equation}
\partial_t \pi^i - \partial_j\sigma^{ij} = 0,\label{Newton}
\end{equation}
which is simply the continuum form of the Newton's equation of motion,
relating the stress imbalance to the rate of change of lattice
momentum.  We will now rewrite the action in terms of fields which solve this constraint explicitly.  First, however, it is convenient to introduce rotated field redefinitions:
\begin{equation}
B^i = \epsilon^{ij}\pi_j,\,\,\,\,\,\,\,\,\,\,\,\,\,\,\,\,\,\,\,E_\sigma^{ij} = \epsilon^{ik}\epsilon^{j\ell}\sigma_{k\ell}.
\end{equation}
(The label $\sigma$ on the field $E_\sigma^{ij}$ is to indicate its relation
to the rotated stress tensor.)  This change of variables transforms the Newton's equation into a generalized Faraday law:
\begin{equation}
\partial_t B^i + \epsilon_{jk}\partial^j E_\sigma^{ki} = 0.
\end{equation}
This specific Faraday equation is precisely that which occurs in the scalar charge tensor gauge theory.  The general solution to this equation is conveniently given by the potential formulation of the gauge theory, in terms of a symmetric rank-2 tensor gauge field, $A^{ij}$, and a scalar potential, $\phi$:
\begin{equation}
  B^i = \epsilon_{jk}\partial^j A^{ki}\,,\,\,\,\,\,\,\,\,\,
E_\sigma^{ij} = -\partial_t A^{ij} - \partial_i\partial_j\phi\;.
\label{potentials}
\end{equation}
in close analogy with the potential formulation of Maxwell theory.  In the static limit, we can use the relation between $\sigma^{ij}$ and $E_{\sigma}^{ij}$ to write:
\begin{equation}
\sigma^{ij} = \epsilon^{ik}\epsilon^{j\ell}\partial_k\partial_\ell\phi,
\end{equation}
demonstrating that $\phi$ plays the role of the Airy stress function of static
elasticity theory.  Note that the fields $E_\sigma^{ij}$ and $B^i$
are invariant under the generalized gauge transformation on the
potentials,
\begin{equation}
A_{ij}\rightarrow A_{ij}+\partial_i\partial_j\alpha\,,
\,\,\,\,\,\,\,\,\,\,\,\,\,\,\,\,\,\phi\rightarrow\phi+\partial_t\alpha,
\end{equation}
for arbitrary function $\alpha(\vec{x},t)$.  The potential formulation has
therefore introduced a gauge redundancy into the problem.  (Importantly, this gauge field is noncompact, as we discuss in more detail in the next section.)  Utilizing these potentials (\ref{potentials}), the action (\ref{action2}) can be written as:
\begin{align}
\begin{split}
  S =& \int
  d^2xdt\bigg(\frac{1}{2}\tilde{C}^{-1}_{ijk\ell}E_\sigma^{ij}E_\sigma^{k\ell}
    - \frac{1}{2}B^iB_i \\
    &+ \epsilon^{ik}\epsilon^{j\ell}\partial_t (A_{k\ell} + \partial_k\partial_\ell \phi)  u_{ij}^{(s)} -\epsilon^{ij}\epsilon_{k\ell}\partial^kA^{\ell j}    \partial_t u_i^{(s)} \bigg),
\end{split}
\end{align}
where $\tilde{C}^{ijk\ell} =
\epsilon^{ia}\epsilon^{jb}\epsilon^{kc}\epsilon^{\ell d}C_{abcd}$ is a rotation of the elastic coefficient tensor.  We can now integrate by parts on the last two terms, being careful that derivatives acting on $u_{ij}^{(s)}$ need not commute, to obtain:
\begin{align}
\begin{split}
  S = \int
  d^2xdt\bigg(\frac{1}{2}\tilde{C}^{-1}_{ijk\ell}E_\sigma^{ij}E_\sigma^{k\ell}&
    - \frac{1}{2}B^iB_i \\
    &+ \rho\phi - J^{ij}A_{ij} \bigg),
\end{split}
\label{finalact}
\end{align}
where we have defined the current tensor $J^{ij}$ as:
\begin{equation}
J^{ij} =
\epsilon^{ik}\epsilon^{j\ell}(\partial_t\partial_k
- \partial_k\partial_t)u_\ell.
\end{equation}
This tensor captures the motion of both dislocations and disclinations, as introduced in Ref.\;\onlinecite{MRvortices,genem}.  For a dislocation with Burgers vector $b^\ell$ at position $x^k(t)$ moving at velocity $v^j$, this tensor takes the form $J^{ij} = \frac{1}{2}(\epsilon^{i\ell}v^{j}b_\ell + \epsilon^{j\ell}v^ib_\ell)\delta^{(2)}(x^k(t))$,\cite{foot5} with the trace $J^i_{\,\,i}$ describing dislocation climb.\cite{MRvortices}

Through the above manipulations, we have successfully mapped the original action of two-dimensional elasticity theory (\ref{orig}) into the action of the scalar-charge tensor gauge theory (\ref{finalact}).  The two polarizations of phonons have mapped onto the two gapless gauge modes of the gauge theory, while the disclinations of the crystal, described by density $\rho$, have mapped onto the fracton charges.  The correspondence between disclinations and fractons becomes particularly clear by examining the Gauss's law of the gauge theory, obtained by varying the action with respect to $\phi$:
\begin{equation}
\partial_i\partial_jE^{ij} = \rho,
\label{gauss}
\end{equation}
where the new electric field tensor $E_{ij}$ (without $\sigma$ subscript) is defined as:
\begin{equation}
E_{ij} = -\partial\mathcal{L}/\partial \dot{A}^{ij} =
\tilde{C}^{-1}_{ijk\ell}E_\sigma^{k\ell}.
\end{equation}
The Gauss's law (\ref{gauss}) serves as the dual formulation of the definition of disclination density (\ref{disc}).  We thereby see that the duality maps $E^{ij}$ to a rotated strain tensor:
\begin{equation}
E^{ij} = \epsilon^{ik}\epsilon^{j\ell}u_{k\ell},
\end{equation}
while we have already seen that the closely-related field $E_\sigma^{ij}$ is mapped to a rotated stress tensor, $E_\sigma^{ij} =
\epsilon^{ik}\epsilon^{j\ell}\sigma_{k\ell}$.  The relation
$E_\sigma^{ij} = \tilde{C}^{ijk\ell}E_{k\ell}$ between the two
electric field tensors exactly mirrors the relation $\sigma^{ij} =
C^{ijk\ell}u_{k\ell}$ between the stress and strain tensors.

\subsection{Instantons and Stability}

In the fracton tensor gauge dual of commensurate crystals, described in the previous section, the gauge-invariant magnetic field operator takes the form:
\begin{equation}
B^i = \epsilon_{jk}\partial^jA^{ki}.
\end{equation}
If we regard $A_{ij}$ as simply a real-valued variable ($i.e.$ noncompact), then there are two separate conservation laws which the magnetic field obeys: conservation of magnetic flux, $\int d^2x\,B^i$, as well as conservation of a first moment of flux, $\int d^2x\,B^ix_i$.  These conservation laws can be derived by rewriting each of these quantities as boundary terms:
\begin{equation}
\int d^2x\,B^i = \int d^2x\,\epsilon_{jk}\partial^jA^{ki} = \oint dn^j \epsilon_{jk}A^{ki},
\end{equation}
\begin{equation}
\int d^2x\,B^ix_i = \int d^2x\,x_i\epsilon_{jk}\partial^jA^{ki} = \oint dn^jx_i\epsilon_{jk}A^{ki}.
\label{magcon}
\end{equation}
Since each of these quantities is encoded on the boundary, no local operator in the bulk of the system can cause them to change, making them locally conserved quantities.  For a noncompact gauge theory such as this, the gapless mode is unambiguously stable, since there are no gauge-invariant mass terms which can be added to the action.  This corresponds to a stable deconfined phase of the gauge theory.  In a \emph{compact} gauge theory, on the other hand, $A^{ij}$ is only defined modulo some compactification radius, say $2\pi$.  In a normal compact $U(1)$ gauge theory, this allows the magnetic flux to slip by $2\pi$.  Such flux slip events, or instantons, destabilize the compact version of Maxwell theory, gapping the photon and confining the charges.\cite{polyakov}  In the present case of a tensor gauge theory, a compact gauge field would imply that the magnetic flux could slip by units of $2\pi$.  Similarly, the moment of flux could slip by units of $2\pi a$, where $a$ is the lattice spacing.  Just as in ordinary $U(1)$ gauge theory, such instantons would destabilize these theories in two dimensions.\cite{alex}

Since the long-range order of a crystal has a finite range of stability in two dimensions, we expect that elasticity theory should map onto the noncompact theory, without the destabilizing instanton processes.  But this fact should naturally arise out of the structure of the duality, without relying on the known stability of elasticity theory.  In order to see where noncompactness arises, it is useful to translate the magnetic conservation laws seen in Equation \ref{magcon} into the original elastic variables, in terms of which we can write:
\begin{equation}
\int d^2x\,\pi^i = \textrm{const.},\,\,\,\,\,\,\,\,\,\,\,\,\,\int d^2x\,\epsilon^{ij}x_i\pi_j = \textrm{const.},
\end{equation}
which correspond to conservation of linear and angular momentum of the crystal.  These two conservation laws follow from the underlying translational and rotational symmetries of space, which importantly are \emph{spontaneously} broken in the crystalline-ordered phase.  Thus, any flux-changing instanton event of the gauge theory maps onto a violation of momentum (or angular momentum) conservation in the elastic theory.  Such terms can arise in specific situations, such as the case of a crystal coupled to an underlying substrate.  But in the absence of a translational and rotational symmetry breaking substrate or external fields, the instantons of the gauge theory are ruled out by the underlying translational and rotational symmetries.

\subsection{Defect Mobility and Continuity Equations}

We have shown that there is a direct correspondence at the operator level between gauge charges of the scalar-charge fracton tensor gauge theory and lattice defects of two-dimensional elasticity theory, with disclinations playing the role of fractons and dislocations acting as dipoles.  However, we have not yet explicitly checked the correspondence between the mobility restrictions on the two sides of the duality, which has a few subtle features.  We now verify that the gauge theory description encodes the expected mobility constraints of elasticity theory.

We begin with the disclinations, corresponding to fracton charges of the gauge theory.  The defining property of fractons is that they cannot move by themselves.  In the scalar charge theory in particular, a charge can only move if it emits or absorbs extra dipoles, in order to keep the total dipole moment fixed.  The emission of such dipoles is energetically costly, so an isolated fracton cannot move without an energy source driving it.  An essentially identical story is known to hold for disclinations.  The motion of a disclination is always accompanied by the creation of dislocation defects, analogous to the creation of dipoles.\cite{emit1,emit2,emit3,emit4}  Since the creation of dislocations costs energy, an isolated disclination is not free to move.  In other words, a disclination is a fracton excitation, as expected from the duality.

One way to phrase this more formally is to study the continuity equation of the theory:
\begin{equation}
\partial_t\rho + \partial_i\partial_jJ^{ij} = 0,
\end{equation}
in various situations.  For example, consider a theory with density $p^i$ of $i$-directed dipoles, moving at velocity $v^j$.  In this case, the charge and current take the form:
\begin{equation}
\rho = \partial_ip^i,\,\,\,\,\,\,\,\,\,\,\,\,J^{ij} = \frac{1}{2}(p^iv^j + v^ip^j).
\end{equation}
Plugging back into the continuity equation, we obtain:
\begin{equation}
\partial_i(\partial_tp^i + p^i\partial_jv^j + v^i\partial_jp^j) = 0.
\end{equation}
Rearranging, we find:
\begin{equation}
\partial_t p^i + p^i\partial_jv^j = -\rho v^i,
\end{equation}
where we have set a constant of integration to zero on physical grounds.  This equation represents a continuity equation for the dipoles of the theory ($p^i$), with a source term corresponding to the motion of fractons ($\rho$).

Now we must address the dipoles and dislocations, which have slightly more subtle behavior.  In a generic tensor gauge theory of the form discussed, there are no further restrictions on particle mobility, and dipoles are fully mobile objects.  The conservation laws of charge and dipole moment do not place any fundamental restriction on the motion of a dipole, so long as its orientation is preserved.  On the other hand, we know that a dislocation defect can only move easily in the direction of its Burgers vector.  Therefore, if the duality is truly to hold, there must be some mechanism which impedes longitudinal motion of dipoles in the gauge theory.  We will see that this mechanism corresponds to the presence of an extra symmetry in the gauge dual of elasticity theory, corresponding to the $U(1)$ symmetry of atom number conservation.  This results in the dipoles of the gauge theory exhibiting the symmetry-enforced mobility restriction of only moving perpendicular to their dipole moment, as we discuss in more detail later.

To see this mechanism at work, we consider a particular component of the quadrupole moment, $Q^i_{\,\,i} = \int d^2x\,\rho x^2$, which changes under longitudinal dipole motion.  Through integration by parts, it is straightforward to obtain the following relation:
\begin{equation}
\int_V d^2x\,(\rho x^2 - 2E^i_{\,\,i}) = \int_{\partial V} dn_i\,(x^2\partial_j E^{ij} - 2x_j E^{ij}).
\end{equation}
The right-hand side of this equation is just a boundary term, which cannot be changed by local operations in the bulk of the region.  In other words, the quantity $(\rho x^2 - 2E^i_{\,\,i})$ obeys a local conservation law.  Any change in $\rho x^2$ is necessarily accompanied by an opposing change in $E^i_{\,\,i}$.  To understand the physical meaning of this conservation law, we can rewrite the trace in elasticity language as:
\begin{equation}
E^i_{\,\,i} = \partial_i u^i,
\end{equation}
which corresponds to volume changes of the lattice.  This indicates that any motion of a dislocation transverse to its Burgers vector must be accompanied by stretching or compressing of the lattice.  In a solid, such compressions are very energetically costly.  In particular, when $\partial_i u^i$ becomes large, it is useful to write:
\begin{equation}
E^i_{\,\,i} = \partial_iu^i = n_d + \partial_i \tilde{u}^i,
\end{equation}
where $n_d$ is the number of vacancies minus the number of interstitial defects, and $\tilde{u}^i$ is a smooth function obeying $\partial_i \tilde{u}^i\ll 1$.  Up to boundary terms, we can then write:
\begin{equation}
\int_V d^2x\,(\rho x^2 - 2n_d) = \textrm{constant}.
\end{equation}
In other words, the longitudinal motion of a dipole ($i.e.$ motion of a dislocation transverse to its Burgers vector) necessarily involves the absorption or creation of vacancies or interstitial defects, which are closely related to the quadrupole moment of the gauge theory.  This is in line with the fact that, in certain fracton models, dipoles can only move in certain directions via the absorption or emission of quadrupolar excitations.  This provides an energetic barrier which makes the dislocations into quasi-one-dimensional particles, as expected.  It is important to note that this one-dimensional behavior is closely tied to the vacancy quantum number and its associated $U(1)$ symmetry.  If, for example, the $U(1)$ symmetry were broken by vacancies forming a condensate, such that vacancies could be easily absorbed and emitted by the dipole, then these mobility restrictions would be lifted, as we will discuss further in a later section.  In this sense, we have a type of symmetry-enriched fracton behavior, with immobility protected by conservation of vacancy number.

We can more formally see that transverse dislocation motion creates vacancy/interstitial defects by examining the Ampere equation of motion, first studied in Reference \onlinecite{MRvortices}, which follows directly from the Hamiltonian:
\begin{equation}
\partial_t E^{ij} + \frac{1}{2}(\epsilon^{ik}\partial_kB^j + \epsilon^{jk}\partial_kB^i) = -J^{ij}.
\end{equation}
The piece of this equation which is relevant for our purposes is the trace, which takes the form:
\begin{equation}
\partial_t E^i_{\,\,i} + \epsilon^{ij}\partial_iB_j = -J^i_{\,\,i}.
\end{equation}
We can rewrite the left hand side in terms of elasticity variables to obtain\cite{MRvortices}:
\begin{equation}
\partial_t n_d + \partial_i\pi^i = -J^i_{\,\,i},
\label{continuity}
\end{equation}
where we have used the fact that $E^i_{\,\,i}\sim n_d$, the number of vacancies minus interstitials, since $\partial_i \tilde{u}^i\ll 1$.  We note that $\pi^i = J^i_{(d)}$ plays the role of the current of vacancies/interstitials.  The above equation then represents a continuity equation for the vacancy number, sourced by $J^i_{\,\,i}$, the trace of the current tensor.  This trace represents the rate of longitudinal motion of dipoles (transverse motion of dislocations).\cite{genem}  Equation \ref{continuity} therefore formally shows that transverse motion of dislocations will create vacancy/interstitial defects, in line with our earlier arguments.

\subsection{Forces, Interactions, and Energetics}

Another important aspect on both sides of the duality is the way in which the charges/defects interact with each other, mediated by long-range fields.  Here we discuss the structure of the interactions between charges on the gauge theory side, which match up perfectly with the expected properties of elasticity theory.  Towards this end, the first important piece of information is the force on charges due to the electric and magnetic fields of the gauge theory.  Since an isolated fracton cannot move, there is no meaningful sense of force on it.  Rather, we should discuss the force on dipoles, which serve as the fundamental mobile objects of the theory.  Indeed, dipoles are the objects which behave most like conventional particles in this theory, with $-p_jA^{ij}$ serving as the effective gauge field seen by dipole $p_j$.\cite{genem}  One way to see this is to consider the form of the matter to gauge field coupling in the action:
\begin{equation}
S_{coup} = \int d^2xdt\,(J^{ij}A_{ij} + \rho\phi),
\end{equation}
where, for a single dipole $p^i$ at instantaneous position $r^i(t)$ and velocity $v^i = \dot{r}^i$, the density and current are given by:
\begin{equation}
\rho = p^i\partial_i\delta(x^k-r^k(t)),
\end{equation}
\begin{equation}
J^{ij} = \frac{1}{2}\bigg(p^i\dot{r}^j\delta(x^k-r^k(t)) + p^j\dot{r}^i\delta(x^k-r^k(t))\bigg).
\end{equation}
Using these forms, it becomes clear that the effective scalar and vector potentials for a dipole $p_j$ are simply given by $-p_jA^{ij}$ and $-p_j\partial^j\phi$, at which point the derivation of the equation of motion proceeds exactly as in the case of ordinary Maxwell theory.  By varying the action $S_{coup}$ with respect to $r^i(t)$, we obtain the equation of motion for a dipole as:
\begin{align}
p_i(\partial^i\partial^j\phi + \partial_tA^{ij}) - p_i\epsilon^{jk}v_k\epsilon_{mn}\partial^mA^{ni} \nonumber\\
= -p_i(E^{ij} + \epsilon^{jk}v_kB^i) = 0.
\end{align}
Assuming an effective mass $m$ description of a dipole will give an additional 
inertial term, allowing us to write:
\begin{equation}
F^j = m\ddot{r}^j = -p_i(E^{ij} + \epsilon^{jk}v_kB^i).
\end{equation}
The above equation serves as the fundamental Lorentz force on dipoles, as described in more detail in Reference \onlinecite{genem}.  As in standard electromagnetism, static charges feel only a force from an electric field, while moving charges experience a velocity-dependent force.  Translating this force into elastic variables, we can write the force on a dislocation $b^i = \epsilon^{ij}p_j$ as:
\begin{equation}
F^j = \epsilon^{jk}b^i(\sigma_{ik} + v_k\pi_i).
\end{equation}
The first term is the standard Peach-Koehler force on a static dislocation\cite{peach}, while the second is a velocity-dependent correction which will be much smaller than the ``electric" contribution, since the typical dislocation velocity will be much smaller than the phonon velocity, which serves as the effective ``speed of light."  For the rest of this section, we will assume that the dislocation velocity is small, such that we can work in the electrostatic limit, writing:
\begin{equation}
F^j = -p_iE^{ij} = \epsilon^{jk}b^i\sigma_{ik}.
\end{equation}
As in conventional electromagnetism, this electrostatic limit can be conveniently treated by introducing a potential formulation, which is a significant simplification of the problem.  We can derive this potential formulation from the static limit of the generalized Faraday's equation:
\begin{equation}
\partial_t B^i = -\epsilon_{jk}\partial^j E^{ki}_\sigma = 0.
\end{equation}
The condition $\epsilon_{jk}\partial^j E^{ki}_\sigma = 0$ has the general solution:
\begin{equation}
E^{ij}_\sigma = \partial^i\partial^j\phi,
\end{equation}
for scalar potential $\phi$, which plays the role of the Airy stress function of elasticity theory.  Plugging $\phi$ into the Gauss's law of the theory, and using the relation $E_{ij} = \tilde{C}_{ijk\ell}^{-1}E^{k\ell}_\sigma$, we obtain a generalized Poisson equation:
\begin{equation}
\tilde{C}^{-1}_{ijk\ell}\partial^i\partial^j\partial^k\partial^\ell  \phi = \rho.
\end{equation}
The physical meaning of $\phi$ can be clarified by writing the electrostatic energy as:
\begin{align}
H &= \frac{1}{2}\int d^2x\,(\tilde{C}^{ijk\ell} E_{ij}E_{k\ell}) \nonumber\\
 &= \frac{1}{2}\int d^2x\,\tilde{C}^{ijk\ell}\tilde{C}^{-1}_{ijnm}\partial^n\partial^m\phi E_{k\ell}  \nonumber\\
 &=\frac{1}{2}\int d^2x\,\rho\phi.
\end{align}
We see that $\phi$ represents the potential energy per unit charge, justifying our use of the term ``potential."  In terms of this potential, the force on a dipole is given by:
\begin{equation}
F^j = -p_i\partial^i\partial^j\phi.
\end{equation}

In order to determine how this force depends on the distance from other crystalline defects, we now find the appropriate potentials for all excitations of the theory.  For simplicity, we focus on a crystal of high symmetry ($e.g.$ a hexagonal crystal) such that only isotropic terms appear up to fourth order in derivatives.  In this case, up to normalization, our Poisson equation takes the form:
\begin{equation}
\partial^4\phi = \rho.
\end{equation}
For an isolated fracton of charge $q$ ($i.e.$ a disclination), the potential must then satisfy:
\begin{equation}
\partial^4\phi_q = q\,\delta^{(2)}(r).
\end{equation}
Simply by dimensional analysis, we can conclude that the generic solution to this equation is:
\begin{equation}
\phi_q(r) = \alpha r^2\log \frac{r}{r_0},
\end{equation}
for some constants $\alpha$ and $r_0$.  One can readily check that this equation only solves the Poisson equation if we choose $\alpha = q/8\pi$.  As discussed in earlier treatments of elasticity theory\cite{boundary,boundary2}, we have a condition of vanishing stress at the boundary which requires $r_0\sim L$.  The final form of the potential is then:
\begin{equation}
\phi_q(r) = \frac{qr^2}{8\pi}\log\frac{r}{L}.
\end{equation}
The growth of the potential as a function of $r$ is indicative of the extensive energy cost necessary to create isolated disclinations within the solid phase.  Separating a group of disclinations out to a distance $R$ in Euclidean space would require an energy of order $R^2$.  In practice, therefore, disclination physics is typically only seen on small scales.  In an appropriate sense, the disclinations are ``confined" in the crystal (though this elastostatic mechanism is different from more conventional forms of confinement\cite{sub}).  Nevertheless, the disclinations will play an important role in the melting transitions of the crystal, particularly in the hexatic phase, so we must account for them in a complete description.  In the vicinity of a disclination, dislocations experience a large force given by:
\begin{align}
F^j &= -p_i\partial^i\partial^j\phi_q(r) \nonumber\\
&= -\frac{qp^j}{8\pi}(1 + 2\log\frac{r}{L}) - \frac{q(p\cdot   r)r^j}{4\pi r^2},
\end{align}
which grows logarithmically with distance.

In addition to the interaction with individual charges/disclinations, we should also determine the interaction between dipoles/dislocations.  Since a dipole is simply a bound state of two fracton charges, we can easily determine the potential generated by a dipole $p^i$ to be:
\begin{align}
\phi_p(r) &= -p^i\partial_i\bigg(\frac{r^2\log (r/L)}{8\pi}\bigg)\nonumber\\
&= -\frac{(p\cdot   r)}{8\pi}(2\log (r/L) + 1).
\end{align}
We can also determine the effect of such a potential on other dipoles.  Since a dipole is electrically neutral, it is only sensitive to the derivative of $\phi$.  As such, the effective potential energy between two dipoles, $p$ and $p'$, is given by:
\begin{align}
V_{pp'}(r) &= p^i\partial_i\phi_{p'} \nonumber\\
&=-\frac{(p\cdot   p')}{8\pi}(2\log (r/L) + 1) - \frac{(p'\cdot   r)(p\cdot   r)}{4\pi r^2}.
\end{align}
For two oppositely directed dipoles, $p' = -p$, the long-distance behavior is a simple logarithmic attractive potential:
\begin{equation}
V_{pp'}(r)\rightarrow \frac{p^2}{4\pi}\log \frac{r}{a},
\label{limit}
\end{equation}
where we have subtracted off the self-energy, $\frac{p^2}{4\pi}\log(L/a)$, associated with two well-separated dipoles.  These expressions agree with those discussed by Halperin and Nelson\cite{halperin}, restricted to the case of structures such as hexagonal crystal, with very high symmetry.  Creating an isolated dislocations costs an energy of order $\log L$, which has important consequences for the melting transitions of solids.  In this limit, the force between dipoles takes the form:
\begin{equation}
F^j = -\partial^jV_{pp'}(r) = -\frac{p^2r^j}{4\pi r^2},
\end{equation}
which is purely radial, and is equivalent to a force between ordinary two-dimensional electric charges.

\subsection{Introducing Matter Fields}

The duality mapping of elasticity theory yielded the following gauge theory action:
\begin{equation}
S = \int d^2xdt\bigg(\frac{1}{2}\tilde{C}_{ijk\ell}^{-1}E^{ij}_\sigma E^{k\ell}_\sigma - \frac{1}{2}B^iB_i - \rho\phi - J^{ij}A_{ij}\bigg),
\end{equation}
with source terms corresponding to fracton charge ($\rho$) and dipole current ($J^{ij}$).  However, this action does not feature separate fields describing the charges, and the dynamics of charges is not made explicit.  We now seek to rewrite the action in a form which manifestly captures the dynamics of charges, as described by charged matter fields.  To this end, we first introduce by hand the core energy and kinetic energy of charges to the action:
\begin{align}
\begin{split}
S = \int d^2xdt\bigg(&\frac{1}{2}\tilde{C}_{ijk\ell}^{-1}E^{ij}_\sigma E^{k\ell}_\sigma - \frac{1}{2}B^iB_i \\
&- \rho\phi - J^{ij}A_{ij} - E_c\rho^2 - g J_{ij}J^{ij}   \bigg).
\end{split}
\label{sourceact}
\end{align}
Such terms arise from short-distance physics, outside the scope of the original linearized elasticity theory, and are necessary for a sensible discussion of charges.  We now go to the path integral formulation of the theory, which integrates not only over all configurations of $A_{ij}$ and $\phi$, but also over all possible configurations of $\rho$ and $J_{ij}$:
\begin{equation}
Z = \int\,\mathcal{D}A_{ij}\mathcal{D}\phi\,\mathcal{D}\rho\mathcal{D}J_{ij}\,e^{iS}.
\end{equation}
The ``integration" over $\rho$ should technically be a sum over only a discrete set of values.  However, replacing this sum by an integral results in only a minor difference in the theory, as we comment on further below.  As Gaussian variables, we can now integrate $\rho$ and $J_{ij}$ out of the path integral to yield:
\begin{align}
\begin{split}
S = \int d^2xdt\bigg(\frac{1}{2}\tilde{C}_{ijk\ell}^{-1}E^{ij}_\sigma E^{k\ell}_\sigma - \frac{1}{2}B^iB_i \\
+ \frac{1}{2}g_0\phi^2 + \frac{1}{2}g_1 A_{ij}A^{ij}\bigg),
\end{split}
\label{fixed}
\end{align}
for constants $g_0$ and $g_1$.  Note that this action does not appear to be gauge-invariant, corresponding to the choice of unitary gauge.  However, we can restore gauge invariance to the entire action by introducing an auxiliary field $\theta$, which couples to the gauge field as:
\begin{align}
\begin{split}
S = \int &d^2xdt\bigg(\frac{1}{2}\tilde{C}_{ijk\ell}^{-1}E^{ij}_\sigma E^{k\ell}_\sigma - \frac{1}{2}B^iB_i \\
&+ \frac{1}{2}g_0(\partial_t\theta - \phi)^2 + \frac{1}{2}g_1 (\partial_i\partial_j\theta - A_{ij})^2\bigg).
\end{split}
\label{supact}
\end{align}
The action is now gauge-invariant under the full set of transformations:
\begin{subequations}
\begin{equation}
A_{ij}\rightarrow A_{ij} + \partial_i\partial_j\alpha,
\end{equation}
\begin{equation}
\phi\rightarrow\phi + \partial_t\alpha,
\end{equation}
\begin{equation}
\theta \rightarrow\theta + \alpha.
\end{equation}
\end{subequations}
The original form of Equation \ref{fixed} can be recovered by gauge-fixing the above action and setting $\theta = 0$.  However, this new form of the action has the advantage of manifest gauge invariance, featuring an explicit field $\theta$ capturing the dynamics of charges.  Note that, had we accounted for discreteness of charge, the only change to this action would be the replacement of $(\partial_t\theta-\phi)^2$ by $\cos(\partial_t\theta -\phi)$, which is important for describing the transition between fracton insulating and superconduting states.  The crystalline phase corresponds to a fracton insulator, where the matter is gapped and the low-energy theory is the pure Maxwell piece of the action in Equation \ref{sourceact}.  In the melted phase, corresponding to the condensed ``superconducting" phase of charges, we can expand the cosine around its minimum, recovering the action in Equation \ref{supact}.  Starting from this ``superconducting" action, we can then access the conventional crystal by condensing the topological defects of the fracton condensate.

\subsection{External Stress}

We have now established a gauge dual formulation of an isolated crystal.  However, it is also useful to consider a crystal subjected to an externally applied stress encoded via a tensor $\Sigma_{ij}$.  In this case, the action will be modified to include a source term for $u_{ij}$ as follows:
\begin{equation}
S = \int d^2xdt\frac{1}{2}\bigg[(\partial_tu^i)^2 - C^{ijk\ell}u_{ij}u_{k\ell} - \Sigma^{ij}u_{ij}\bigg].
\end{equation}
We can once again introduce Hubbard-Stratonovich fields to transform the action into:
\begin{align}
S = \int d^2xdt\bigg[\frac{1}{2}&C^{-1}_{ijk\ell}\sigma^{ij}\sigma^{k\ell} - \frac{1}{2}\pi^i\pi_i \nonumber\\
&- (\sigma^{ij} + \Sigma^{ij})u_{ij} + \pi^i\partial_t u_i\bigg].
\end{align}
As before, we can decompose $u_i$ into its smooth and singular pieces.  Upon integrating out the smooth piece, we obtain the following constraint:
\begin{equation}
\partial_t u^i -\partial_j(\sigma^{ij} + \Sigma^{ij}) = 0.
\end{equation}
To explicitly solve this equation, we introduce rotated field redefinitions:
\begin{equation}
B^i = \epsilon^{ij}\pi_j,\,\,\,\,\,\,\,\,\,\,\,\,\,E^{ij}_\sigma = \epsilon^{ij}\epsilon^{j\ell}(\sigma_{ij} + \Sigma_{ij}).
\end{equation}
We can then represent these rotated fields in terms of the usual potential formulation:
\begin{equation}
B^i = \epsilon_{jk}\partial^jA^{ki},\,\,\,\,\,\,\,\,\,\,\,\,\,\,\,E^{ij}_\sigma = -\partial_tA^{ij} - \partial^i\partial^j\phi.
\end{equation}
In the gauge dual language, the elastic action subject to external stress can be written as:
\begin{align}
S = \int d^2xdt\bigg[\frac{1}{2}\tilde{C}^{-1}_{ijk\ell}E_\sigma^{ij}E_\sigma^{k\ell} - \tilde{C}^{-1}_{ijk\ell}E^{ij}_\sigma\tilde{E}_\sigma^{k\ell} \nonumber\\
 - \frac{1}{2}\pi^i\pi_i - J^{ij}A_{ij} - \rho\phi\bigg],
\end{align}
where we have defined $\tilde{E}^{ij}_\sigma = \epsilon^{ik}\epsilon^{j\ell}\Sigma_{k\ell}$, and have dropped an overall constant term which is quadratic in $\tilde{E}^{ij}_\sigma$.  All other quantities are the same as in the unstressed case.  In this language, we can see that the gauge dual for a crystal subject to an applied external stress is a tensor gauge theory subject to an applied external electric tensor field $\tilde{E}^{ij}_\sigma$.  This tensor electric field will exert a force on dipoles in the dual description, which corresponds to the fact that an external stress exerts a force on dislocations in a crystal.

\subsection{An Alternative Formulation of Elasticity Theory}
\label{sec:alt}

In the preceding sections, we have formulated the low-energy elasticity theory of crystals purely in terms of the symmetric strain tensor, $u_{ij} = \frac{1}{2}(\partial_iu_j + \partial_j u_i)$, as the antisymmetric part corresponds to the bond angle $\theta = \frac{1}{2}\epsilon^{ij}\partial_i u_j$, which is forbidden to appear by itself by the underlying rotational invariance of the crystal.  Below, we use a reformulated rotationally invariant version of elasticity theory in terms of an unsymmetrized strain tensor, but with the bond angle $\theta$ appearing to ensure overall rotational invariance and equivalence to the conventional formulation.  This new formulation also has interesting implications for fracton physics, as discussed more fully in Reference \onlinecite{vectorfracton}.  In this formulation, we start from an action featuring all spatial derivatives of $u_i$, both symmetric and antisymmetric, plus an angular variable $\theta$ representing the local orientation of the crystal, which $a$ $priori$ we allow to be independent of $u_i$.  Importantly, however, we demand that the action of the theory be invariant under shifting the local orientation $\theta$ and the bond angle $\frac{1}{2}\epsilon^{ij}\partial_iu_j$ by equal amounts.  Consistent with this restriction, the most general low-energy action to linear order takes the form:
\begin{align}
S = \int d^2x&dt\frac{1}{2}\bigg[(\partial_t u^i)^2 + (\partial_t\theta)^2 - (\partial_i\theta)^2 \nonumber\\
&- C^{ijk\ell}(\partial_iu_j - \epsilon_{ij}\theta)(\partial_ku_\ell - \epsilon_{k\ell}\theta) \bigg].
\label{altelast}
\end{align}
Written in this way, we see that the action for the antisymmetric strain tensor is effectively of the form seen in the context of the Higgs mechanism, at low energies Higgsing out the antisymmetric part of $\partial_iu_j$ and thereby reducing it to the conventional formulation in terms of symmetrized strain $u_{ij}$ only.  As in the case of a gauge field acquiring mass through the Higgs mechanism, the field $\theta$ is ``eaten" by the antisymmetric strain, which thereby is eliminated from the low-energy gapless sector of the theory.  In this way, $\theta$ and $\epsilon^{ij}\partial_iu_j$ mutually remove each other from the effective action of the theory, which reduces to the symmetric strain formalism of Equation \ref{elastact}.

Now that we have an alternative action for the theory of elasticity, we can construct a dual gauge theory in much the same way as before.  Introducing Hubbard-Stratonovich fields $\sigma_{ij}$, $\pi_i$, $j_i$, and $L$, we can rewrite the action of Equation \ref{altelast} in the following form:
\begin{align}
S = \int d^2x&dt\frac{1}{2}\bigg[C^{-1}_{ijk\ell}\sigma^{ij}\sigma^{k\ell} - \pi^i\pi_i + j^ij_i - L^2 \nonumber\\
&- \sigma^{ij}(\partial_iu_j - \epsilon_{ij}\theta) + \pi^i\partial_t u_i - j^i\partial_i\theta + L\partial_t\theta  \bigg].
\end{align}
Note that the field $\sigma_{ij}$, playing the role of the stress tensor, is no longer manifestly symmetric.  The field $L$ represents the local angular momentum of the crystal, while $j^i$ represents a current of this angular momentum.  It is now useful to break up both $u_i$ and $\theta$ into their smooth single-valued pieces (denoted by tildes) and their singular pieces, which serve as sources for topological defects:
\begin{equation}
u_i = \tilde{u}_i + u^{(s)}_i,
\end{equation}
\begin{equation}
\theta = \tilde{\theta} + \theta^{(s)}.
\end{equation}
Integrating over the smooth pieces, our action becomes:
\begin{align}
&S = \int d^2xdt\frac{1}{2}\bigg[C^{-1}_{ijk\ell}\sigma^{ij}\sigma^{k\ell} - \pi^i\pi_i + j^ij_i - L^2 \nonumber\\
- \sigma^{ij}&(\partial_iu^{(s)}_j - \epsilon_{ij}\theta^{(s)}) + \pi^i\partial_t u^{(s)}_i - j^i\partial_i\theta^{(s)} + L\partial_t\theta^{(s)}  \bigg],
\label{sepact}
\end{align}
subject to two additional constraints:
\begin{equation}
\partial_t\pi_j - \partial^i\sigma_{ij} = 0,
\end{equation}
\begin{equation}
\partial_t L + \partial_i j^i - \epsilon^{ij}\sigma_{ij} = 0.
\end{equation}
The first equation represents the Newton's equation of motion, relating forces to change in momentum, while the second relates torques to changes of angular momentum.  We now seek to solve these equations explicitly through a potential formulation.  We begin by introducing field redefinitions as follows:
\begin{equation}
B^i = \epsilon^{ij}\pi_j,\,\,\,\,\,\,\,\,\,\,\,\,\,E_\sigma^{ij} = -\epsilon^{ik}\epsilon^{j\ell}\sigma_{k\ell},\nonumber
\end{equation}
\begin{equation}
b = L,\,\,\,\,\,\,\,\,\,\,\,\,\,\,\,\,\,\,\,\,\,\,\,e^i = \epsilon^{ij}j_j,
\end{equation}
in terms of which the constraint equations take the form of generalized Faraday equations:
\begin{equation}
\partial_t B^i + \epsilon_{jk}\partial^j E_\sigma^{ki} = 0,
\end{equation}
\begin{equation}
\partial_tb + \epsilon_{ij}\partial^ie^j + \epsilon_{ij}E_\sigma^{ij} = 0.
\end{equation}
These equations are exactly solved by the following potential formulation:
\begin{subequations}
\begin{equation}
E_\sigma^{ij} = -\partial_tA^{ij} + \partial^i\lambda^j,
\end{equation}
\begin{equation}
B^i = \epsilon_{jk}\partial^jA^{ki},
\end{equation}
\begin{equation}
e^i = -\partial_ta^i - \partial^i\phi - \lambda^i,
\end{equation}
\begin{equation}
b = \epsilon_{ij}(\partial^ia^j - A^{ij}),
\end{equation}
\end{subequations}
where $A^{ij}$ is an arbitrary tensor, without any symmetry properties.  Note that the electric and magnetic fields are invariant under the following transformation on the gauge fields:
\begin{subequations}
\begin{equation}
A_{ij}\rightarrow A_{ij} + \partial_i\alpha_j,
\end{equation}
\begin{equation}
\lambda_i\rightarrow\lambda_i + \partial_t\alpha_i,
\end{equation}
\begin{equation}
a_i\rightarrow a_i + \alpha_i + \partial_i\beta,
\end{equation}
\begin{equation}
\phi\rightarrow\phi + \partial_t\beta,
\end{equation}
\end{subequations}
for two arbitrary gauge functions $\alpha_i(x)$ and $\beta(x)$.  In terms of these new fields, we can rewrite the action from Equation \ref{sepact} as:
\begin{align}
S = \int d^2xdt\frac{1}{2}\bigg[\tilde{C}^{-1}_{ijk\ell}E_\sigma^{ij}E_\sigma^{k\ell} - B^iB_i + e^ie_i - b^2 \nonumber\\
+(\partial_tA^{ij} - \partial^i\lambda^j)(\epsilon_{ik}\epsilon_{j\ell}\partial^ku^\ell_{(s)} - \epsilon_{ij}\theta^{(s)})\nonumber \\
+ \epsilon_{ij}(\epsilon_{\ell k}\partial^\ell A^{ki})\partial_t u_{(s)}^j + \epsilon_{ij}(\partial_ta^i + \partial^i\phi + \lambda^i)\partial_j\theta^{(s)}\nonumber\\
 + (\epsilon_{ij}(\partial^ia^j - A^{ij}))\partial_t\theta^{(s)}  \bigg].
\end{align}
After a few integrations by parts, we can convert the last few terms into source terms for the gauge fields, as follows:
\begin{align}
S = \int d^2xdt\frac{1}{2}\bigg[\tilde{C}^{-1}_{ijk\ell}E_\sigma^{ij}E_\sigma^{k\ell} - B^iB_i + e^ie_i - b^2 \nonumber\\
+A_{ij}J^{ij} + \lambda^jp_j + a_ij^i - \phi s  \bigg],
\label{altdual}
\end{align}
where we have defined the charge and current densities in terms of commutators of derivatives on the singular parts of the fields:
\begin{subequations}
\begin{equation}
J^{ij} = \epsilon^{ik}\epsilon^{j\ell}(\partial_k\partial_t - \partial_t\partial_k)u_\ell^{(s)},
\end{equation}
\begin{equation}
p^j = \epsilon^{ik}\epsilon^{j\ell}\partial_i\partial_ku_\ell^{(s)},
\end{equation}
\begin{equation}
j^i = \epsilon^{ij}(\partial_j\partial_t - \partial_j\partial_t)\theta^{(s)},
\end{equation}
\begin{equation}
s = \epsilon_{ij}\partial^i\partial^j\theta^{(s)}.
\end{equation}
\end{subequations}
The source fields $J^{ij}$ and $p^j$ physically correspond to the current and charge density of dislocations, $i.e.$ point defects around which the lattice displacement $u_i$ has nontrivial winding.  (More accurately, $p^j$ represents the dipole density, which is a simple rotation of the dislocation density, $p_i = \epsilon_{ij}b^j$.)  Concomitantly, $j^i$ and $s$ represent the current and charge density of disclinations, $i.e.$ point defects around which the bond angle $\theta$ has nontrivial winding.  It is also instructive to see how these charges enter the Gauss's laws of the theory, which can be obtained by integrating the Lagrange multipliers $\lambda_i$ and $\phi$ out of the theory, yielding:
\begin{equation}
\partial_iE^{ij} - e^j = p^j,
\end{equation}
\begin{equation}
\partial_ie^i = s.
\end{equation}
By taking a divergence of the first equation and plugging in the second, we obtain:
\begin{equation}
\partial_i\partial_jE^{ij} = s + \partial_ip^i,
\end{equation}
which reflects the fact that the total disclination density, $\partial_i\partial_jE^{ij}$, has contributions both from bare disclinations, $s$, and the dislocations (dipoles) of the system.

In Equation \ref{altdual}, we now have an alternative formulation of the theory of elasticity phrased in terms of a nonsymmetric tensor gauge field $A_{ij}$ and a conventional vector gauge field $a_i$.  To see how this formalism reduces to the previous formulation in terms of symmetric tensors, it is useful to write out the Maxwell portion of the action more explicitly in terms of the potentials for the vector gauge field:
\begin{align}
S_{Max} = \int d^2xdt\frac{1}{2}\bigg[\tilde{C}^{-1}_{ijk\ell}E_\sigma^{ij}E_\sigma^{k\ell} - B^iB_i&\nonumber\\
- (\epsilon_{ij}\partial^ia^j - \epsilon_{ij}A^{ij})^2& \nonumber\\
+ (\partial_ta^i + \partial^i\phi + \lambda^i)(\partial_ta_i& + \partial_i\phi + \lambda_i) \bigg].
\end{align}
Written in this way, it becomes apparent that the action is that of a Higgsed phase for the antisymmetric tensor field $\epsilon_{ij}A^{ij}$, with the curl of the vector gauge field, $\epsilon_{ij}\partial^ia^j$, acting as the phase field of a condensate.  In this way, the antisymmetric component of $A_{ij}$ is gapped out via ``eating" the curl of $a_i$.  Simultaneously, the remaining curl-free component of $a_i$ features in the last term as a phase field gapping the curl of $\lambda^i$ out of the low-energy theory.  Within the low-energy sector, we can then write $\lambda_i = \partial_iA_0$, reducing to the potential of the tensor gauge sector to a simple scalar, as in our earlier duality.  In this way, the vector gauge sector is entirely eaten by the tensor gauge sector, thereby imposing a symmetry condition on the tensor gauge field and reducing to our previous analysis.

While the alternative reformulation of elasticity theory described in this section is useful for making the absence of the antisymmetric strain in the low-energy sector more explicit, its dual description also has important implications for fracton tensor gauge theories.  A nonsymmetric tensor gauge theory features charges which generically do not have any restrictions on their mobility.  However, we have now seen how coupling such a nonsymmetric tensor gauge theory to a conventional vector gauge field can enforce symmetry on the tensor and impose mobility restrictions on the charges.  This provides a novel mechanism for driving phase transitions between fracton and non-fracton phases, which will provide an interesting topic of future investigation.  More details on this new vector reformulation of fracton physics can be found in Reference \onlinecite{vectorfracton}.

\section{Generalized Bosonic Crystal Duality}
\label{sec:comp}

In the previous sections, we have seen how the theory of elasticity for an ordinary commensurate two-dimensional solid maps onto a fracton tensor gauge theory, in which mobility restrictions are closely tied to the quantum numbers of the underlying atoms.  However, this treatment did not incorporate the dynamics or statistics of the underlying atoms, as manifested in vacancy defects of the crystalline order.  As such, the previously discussed pure tensor gauge theory is not equipped to describe zero-temperature melting transitions, driven by quantum fluctuations.  While the commensurate crystal was not particularly sensitive to the statistics of the underlying atoms, a sensible description of quantum fluid phases must take these statistics into account.  For example, we expect that fully quantum melting a crystal of bosonic atoms will result in a superfluid phase, as opposed to some completely featureless state.  Indeed, a truly featureless fully gapped phase preserving all symmetries should be impossible in a continuum, as dictated by the Lieb-Schultz-Mattis theorem.

In order to rule out such unphysical phases and to obtain a sensible description of melting transitions between crystalline and superfluid phases, it is necessary to construct a generalized theory which simultaneously treats both types of ordering.  To this end, we first describe the field theory description of a supersolid, in which both types of ordering are present, with nontrivial coupling between the two sectors.  We then perform a duality transformation to construct a gauge theory capturing the properties of the supersolid phase.  The other phases of boson systems, including commensurate crystals and superfluids, can then be obtained from the supersolid through various condensation transitions.

\subsection{Field Theory Description of Supersolids}

A supersolid is a phase of matter featuring both crystalline and superfluid orders, corresponding to spontaneously broken spatial and $U(1)$ symmetries.\cite{Andreev,MosesChen,ketterle}  The simplest physical picture for such a phase is to consider a solid in which vacancy/interstitial defects have condensed, thereby allowing a condensate of the underlying atoms to coexist with the crystalline order.  In order to describe such a phase in field theory language, we must account for fluctuations around both order parameters.  For the crystalline sector, the appropriate variable to use is the lattice displacement field, $u_i(x)$, which describes the fluctuations of atoms around their equilibrium positions.  For the superfluid sector, the low-energy fluctuations can be described in terms of the phase $\phi(x)$ of the condensate.  More formally, we can obtain these variables starting from a bosonic field $\hat{\psi}(x)$ as:
\begin{equation}
\hat{\psi}(x) = \hat{\psi}_0 + \sum_{\textbf{G}}\hat{\psi}_\textbf{G} e^{i\textbf{G}\cdot   \textbf{x}},
\end{equation}
in terms of its long wavelength component, $\hat\psi_0 = \sqrt{\hat n_0} e^{i\hat\phi}$, and reciprocal lattice ($\textbf{G}$) components, $\hat\psi_\textbf{G} = \sqrt{\hat n_\textbf{G}} e^{i\hat\phi+i\textbf{G}\cdot  \hat{\bf u}}$.  The phase variables $\phi$ and $u_i$ are sufficient for describing the low-energy dynamics of the supersolid phase, while the amplitude variables correspond to gapped modes.

In terms of the Goldstone mode fields $u_i$ and $\phi$, the most general low-energy Hamiltonian we can write down, to lowest order in derivatives, takes the form:
\begin{eqnarray}
\hat{\cal H} &=& \frac{1}{2}\rho^{-1}\hat\pi^2
+\frac{1}{2}\tilde C^{ijk\ell}\hat u_{ij}\hat u_{k\ell}
+\frac{1}{2}\tilde K(\nabla\hat\phi)^2
+\frac{1}{2}\chi^{-1}\hat n^2\nonumber\\
&& -\mu\hat n + \tilde{g}_1 \nabla\hat\phi\cdot  \hat\vec{\pi} + \tilde{g}_2 \hat n \hat u_{ii}\ ,
\label{supham}
\end{eqnarray}
where $u_{ij} = \frac{1}{2}(\partial_iu_j + \partial_ju_i)$ is the symmetric strain tensor.  (Note that the corresponding antisymmetric strain tensor cannot appear explicitly in the action to lowest order, due to the underlying rotational symmetry of the system, which is spontaneously broken by the crystalline order.)  The conjugate fields $\hat{\pi}$ and $\hat{n}=\hat{n}_0 + \sum_\textbf{G}\hat{n}_\textbf{G}$ are the momentum and number density, $\mu$ the chemical potential, $\rho$ the boson average mass density, $\tilde{K}$ the superfluid stiffness, $\chi$ the compressibility, and $\tilde{C}^{ijk\ell}$ the tensor of elastic coefficients.  The first five terms of this Hamiltonian represents the standard Hamiltonians for decoupled elastic and superfluid theories, while the final two terms are the lowest-order symmetry-allowed couplings between the two sectors, a current-current and density-density interaction, respectively.

For obtaining the dual gauge theory, it will be useful to first switch to a path integral representation, given by $Z = \int [d\pi][d{\bf u}][d n]\mathcal{D}\phi\,  e^{i S}$, where the action corresponds to $S =\int_{x,t}\left[\pi\cdot  \partial_t{\bf u} - n\partial_t\phi -  \mathcal{H}[\pi,{\bf u}, n, \phi]\right]$, (with $\int_{x,t}\equiv\int d^2x dt$, $\hbar = 1$).  Using the Hamiltonian from Equation \ref{supham}, we can write the action as:
\begin{eqnarray}
S =\int_{x,t}\bigg[\frac{1}{2}\rho(\partial_t{\bf u})^2 
- \frac{1}{2} C^{ijk\ell} u_{ij} u_{k\ell}
+\frac{1}{2}\chi (\partial_t\varphi)^2 \nonumber \\
- \frac{1}{2} K (\nabla\varphi)^2 - g_1\partial_t{\bf u}\cdot  \nabla\varphi
+ g_2\partial_t\varphi\nabla\cdot  {\bf u}\bigg],\,\,\,
\label{St}
\end{eqnarray}
where we have defined a shifted phase field $\varphi = \phi - \mu t$, stiffnesses $K = \tilde K - \rho \tilde{g}_1^2$ and $C_{ijk\ell}=\tilde C_{ijk\ell} - \chi \tilde{g}_2^2\delta_{ij}\delta_{k\ell}$, and couplings $g_1 = \tilde{g}_1\rho$ and $g_2 = \tilde{g}_2\chi$.  The above action will serve as a starting point for deriving the dual gauge theory.

First, however, it is instructive to examine the equations of motion of this theory, which shed further light on the physical interpretation of the $g_1$ and $g_2$ cross terms.  By varying the action with respect to $\varphi$, we obtain the following equation of motion:
\begin{equation}
\partial_t(-\chi\partial_t\varphi - g_2\partial_iu^i) + \partial_i(K\partial^i\varphi + g_1\partial_t u^i) = 0.
\end{equation}
By identifying the total boson number $n$ and current $j^i$ as:
\begin{equation}
n = -\chi\partial_t\varphi - g_2\partial_iu^i,
\end{equation}
\begin{equation}
j^i = K\partial^i\varphi + g_1\partial_t u^i,
\end{equation}
we can regard the equation of motion as the continuity equation for bosons:
\begin{equation}
\partial_t n + \partial_ij^i = 0.
\end{equation}
It is also useful to separately identify the contribution to boson number and current coming from vacancy and interstitial defects:
\begin{equation}
n_d = -\chi \partial_t\varphi,
\end{equation}
\begin{equation}
j_d^i = K\partial^i\varphi.
\end{equation}
In terms of these fields, we can rearrange the equation of motion for $\varphi$ as:
\begin{equation}
\partial_tn_d + \partial_i j^i_d = g_2\partial_t\partial_iu^i -g_1\partial_i\partial_t u^i \equiv J_s,
\end{equation}
where the source term $J_s$ represents the non-conservation of the net vacancy/interstitial defect number.  From the elastic side of the duality, we know that the source term for vacancy/interstitial creation should be proportional to the transverse motion of dislocations, $i.e.$ longitudinal motion of dipoles, as captured by the trace of the tensor current described in previous sections, $J^i_{\,\,i} = \partial_t\partial_i u^i - \partial_i\partial_t u^i$.  Based on physical grounds, we can therefore conclude that $g_1 = g_2$, which guarantees that vacancy/interstitial number is conserved in the absence of topological defects, such that $\partial_i\partial_t u^i = \partial_t\partial_i u^i$.

It is equally informative to consider the equation of motion for the lattice displacement field.  By varying the action of Equation \ref{St} with respect to $u^i$, we obtain the equation of motion as:
\begin{equation}
\partial_t(\rho\partial_tu_i - g_2 \partial_i\varphi) - \partial_j(C^{ijk\ell}u_{k\ell} - g_1\delta^{ij}\partial_t\varphi) = 0.
\end{equation}
By identifying the total momentum and stress of the system as:
\begin{equation}
\pi_{(tot)}^i = \rho\partial_tu^i - g_2\partial^i\varphi,
\end{equation}
\begin{equation}
\sigma_{(tot)}^{ij} = C^{ijk\ell}u_{k\ell} - g_1\delta^{ij}\partial_t\varphi,
\end{equation}
we can write this equation of motion simply as the Newton's force law for the system:
\begin{equation}
\partial_t\pi^i_{(tot)} - \partial_j\sigma_{(tot)}^{ij} = 0.
\end{equation}
In terms of the conventional momentum and strain of the crystal, $\pi^i = \rho\partial_tu^i$ and $\sigma^{ij} = C^{ijk\ell}u_{k\ell}$ respectively, we can also write the equation of motion as:
\begin{equation}
\partial_t\pi^i - \partial_j\sigma^{ij} = g_2\partial_t\partial_i\varphi - g_1\partial_i\partial_t\varphi.
\end{equation}
In the absence of vortex motion in the superfluid sector, the momentum of the condensate and crystal should be conserved separately.  This indicates that, when $\partial_i\partial_t\varphi = \partial_t\partial_i\varphi$, the right-hand side of the above equation should vanish, which once again allows us to conclude that $g_1 = g_2$, on independent physical grounds.  We therefore set $g_1 = g_2\equiv g$ for the rest of this section.  In terms of this parameter, we can write the equation of motion for $u^i$ as:
\begin{equation}
\partial_t\pi^i - \partial_j\sigma^{ij} = g(\partial^i\partial_t - \partial_t\partial^i)\varphi = g\epsilon^{ij}j^{(v)}_j,
\end{equation}
where $j^{(v)}_j$ is the current of vortices.  This equation reflects the physical fact that motion of vortices relaxes supercurrents, thereby transferring momentum from the condensate to the crystal.

\subsection{Hybrid Gauge Dual of a Supersolid}

Now that we understand the field theoretic description of a supersolid, we can construct a gauge dual through a prescription similar to the previous duality derivation.  We first introduce Hubbard-Stratonovich fields $n$, $\pi_i$, $\sigma_{ij}$, and $j_i$, which allow us to rewrite the action of Equation \ref{St} as:
\begin{align}
S =\int_{x,t}\bigg[\pi^i\dot{u}_i - \frac{1}{2}\overline{\rho}^{-1}\pi^2 + \frac{1}{2} \overline{C}^{-1}_{ijk\ell} \sigma^{ij} \sigma^{k\ell} - \sigma_{ij}u^{ij} - n\dot{\varphi} \nonumber \\
-\frac{1}{2}\overline{\chi}^{-1} n^2 + \frac{1}{2} \overline{K}^{-1}j^2 - j^i\partial_i\varphi - \overline{g}\pi^ij_i - \underline{g}C^{-1}_{iik\ell}\sigma^{k\ell}n\bigg].
\label{hub}
\end{align}
The coefficients are chosen such that the original action is obtained upon integrating out the new fields.  Specifically, we have:
\begin{subequations}
\begin{eqnarray}
&\bar K^{-1} = K^{-1} - \rho g^2 K^{-2},\\
&\bar C_{ijk\ell}^{-1} = C_{ijk\ell}^{-1} - \bar\chi g^2 C_{s s i j}^{-1} C_{t t k\ell}^{-1},\\
&\bar\rho = \rho + K^{-1}\rho^2 g^2,\\
&\bar\chi = \chi + \chi^2 g^2 C_{i i j j}^{-1},\\
&\bar g = g/\rho K,\\
&\underline{g} = g\chi^{-1}.
\end{eqnarray}
\end{subequations}
As before, we now break up both $u_i$ and $\varphi$ into smooth ($\tilde{u}_i$, $\tilde{\varphi}$) and singular ($u_i^{(s)}$, $\phi^{(s)}$) pieces, where the smooth pieces are single-valued while the static singular pieces host topological defects.  Note that the action depends linearly on both $u_i$ and $\varphi$, which allows us to integrate the smooth pieces of both variables, imposing the following constraints:
\begin{subequations}
\begin{equation}
\partial_tn + \partial_ij^i = 0,
\end{equation}
\begin{equation}
\partial_t\pi^i - \partial_j\sigma^{ij} = 0,
\end{equation}
\end{subequations}
which are precisely the continuity equation for total boson number and Newton's equation for total momentum, as discussed earlier.

In order to find the general solution to these constraint equations, it is useful to first introduce ``rotated" field redefinitions as follows:
\begin{align}
\sigma_{ij} =& -\epsilon_{ik}\epsilon_{j\ell}E^{k\ell}_{\sigma},\,\,\,\,\,\,\,\,\,\,\,\,\,\,\,\,\,\,\,\pi^i = \epsilon^{ij}B_j,\nonumber\\
&j^i = \epsilon^{ij}e_j,\,\,\,\,\,\,\,\,\,\,\,\,\,\,\,\,\,\,\,\,\,\,\,\,\,\,\,\,\,n=b.
\end{align}
In terms of these new fields, the continuity and Newton's equations take the form of generalized Faraday equations:
\begin{subequations}
\begin{equation}
\partial_tB^i + \epsilon_{jk}\partial^j E_\sigma^{ki} = 0,
\label{far1}
\end{equation}
\begin{equation}
\partial_tb + \epsilon_{jk}\partial^je^k = 0.
\label{far2}
\end{equation}
\end{subequations}
Just as in ordinary electromagnetism, the general solution to these equations can be found by introducing a potential formulation as follows:
\begin{align}
B^i& = \epsilon_{jk}\partial^jA^{ki},\,\,\,\,\,\,\,\,\,\,\,\,E_\sigma^{ij} = -\partial_tA^{ij} -\partial_i\partial_jA_0,\nonumber\\
&b = \epsilon^{ij}\partial_i a_j,\,\,\,\,\,\,\,\,\,\,\,\,\,\,\,\,\,\,\,\,\,\,\,\,\,e^i = -\partial_ta^i - \partial^ia_0,
\end{align}
where $A^{ij}$ is a symmetric tensor gauge field.  Note that the electric and magnetic fields are invariant under the following gauge transformation on the potentials:
\begin{subequations}
\begin{eqnarray}
&A_{ij}\rightarrow A_{ij} + \partial_i\partial_j\alpha,
\label{trans1}\\
&A_0\rightarrow A_0 + \partial_t\alpha,
\label{trans2}\\
&a_i\rightarrow a_i + \partial_i\beta,
\label{trans3}\\
&a_0\rightarrow a_0 + \partial_t\beta,
\label{trans4}
\end{eqnarray}
\end{subequations}
for two independent gauge parameters $\alpha(x)$ and $\beta(x)$ with arbitrary spatial dependence.  In terms of these new fields, we can write the action of Equation \ref{hub} as:
\begin{align}
S =\int_{x,t}\bigg[\frac{1}{2} \hat{C}_{ijk\ell} E^{ij} E^{k\ell} - \frac{1}{2}\overline{\rho}^{-1}B^2 + \frac{1}{2} \overline{K}^{-1}e^2 -\frac{1}{2}\overline{\chi}^{-1} b^2 \nonumber \\
- \overline{g}B^ie_i - \underline{g}E^i_{\,i}b \nonumber\\
+ \epsilon^{ij}\epsilon^{k\ell}\partial_kA_{\ell i}\partial_t u^{(s)}_j + \epsilon^{ik}\epsilon^{j\ell}(\partial_tA_{k\ell} + \partial_k\partial_\ell A_0)u^{(s)}_{ij}\nonumber\\
- \epsilon^{ij}\partial_ia_j\partial_t\varphi^{(s)} + \epsilon^{ij}(\partial_ta_i + \partial_ia_0)\partial_j\varphi^{(s)} \bigg].
\end{align}
After integrating by parts, we can convert the last two lines into source terms for the gauge fields, yielding the final dual gauge theory action:
\begin{align}
S =\int_{x,t}\bigg[\frac{1}{2} \hat{C}_{ijk\ell} E^{ij} E^{k\ell} - \frac{1}{2}\overline{\rho}^{-1}B^2 + \frac{1}{2} \overline{K}^{-1}e^2 -\frac{1}{2}\overline{\chi}^{-1} b^2 \nonumber \\
- \overline{g}B^ie_i - \underline{g}E^i_{\,i}b - J_s^{ij}A_{ij} - sA_0 - j_v^ia_i - n_va_0 \bigg],
\label{supfindual}
\end{align}
where the charge and current densities of disclinations ($s$) and vortices ($v$) are given by:
\begin{subequations}
\begin{equation}
J^{ij}_s = \epsilon^{ik}\epsilon^{j\ell}(\partial_k\partial_t - \partial_t\partial_k)u_\ell^{(s)},
\end{equation}
\begin{equation}
s = \epsilon^{ik}\epsilon^{j\ell}\partial_i\partial_j u_{k\ell}^{(s)},
\end{equation}
\begin{equation}
j^i_v = \epsilon^{ij}(\partial_t\partial_j - \partial_j\partial_t)\varphi^{(s)},
\end{equation}
\begin{equation}
n_v = \epsilon^{ij}\partial_i\partial_j \varphi^{(s)}.
\end{equation}
\end{subequations}
As before, disclinations map onto the fracton charges of a scalar-charge tensor gauge theory, while vortices of the condensate map onto charges of a conventional vector gauge theory.

We can also explicitly introduce matter fields describing the dynamics of charges.  To this end, we introduce several new terms to the action as follows:
\begin{align}
S =\int_{x,t}\bigg[\frac{1}{2} \hat{C}_{ijk\ell} E^{ij} E^{k\ell} - \frac{1}{2}\overline{\rho}^{-1}B^2 + \frac{1}{2} \overline{K}^{-1}e^2 -\frac{1}{2}\overline{\chi}^{-1} b^2 \nonumber \\
- \overline{g}B^ie_i - \underline{g}E^i_{\,i}b - J_s^{ij}A_{ij} - sA_0 - j_v^ia_i - n_va_0 \nonumber \\
-E_{c,s}s^2 - \lambda_s J^{ij}J_{ij} - E_{c,v}n_v^2 - \lambda_v j^i j_i\bigg],
\end{align}
where the terms in the final line correspond to the core energies and kinetic energies of vortices and lattice defects, respectively, which arise from short-distance physics outside of our original proposed harmonic long wavelength description.  As we did in the case of the pure tensor gauge theory, we now integrate $n_v$, $s$, $j_i$, and $J_{ij}$ out of the path integral to obtain an action as:
\begin{align}
S =\int_{x,t}\bigg[\frac{1}{2} \hat{C}_{ijk\ell} E^{ij} E^{k\ell} - \frac{1}{2}\overline{\rho}^{-1}B^2 + \frac{1}{2} \overline{K}^{-1}e^2 -\frac{1}{2}\overline{\chi}^{-1} b^2 \nonumber \\
- \overline{g}B^ie_i - \underline{g}E^i_{\,i}b + \frac{1}{2}(c_1 A_0^2 +c_2 A_{ij}A^{ij} + c_3 a_0^2 + c_4 a^ia_i) \bigg].
\end{align}
Once again, we have treated the densities and currents as real-valued, as opposed to quantized quantities, a point to which we return later.  Note that the action is now no longer invariant under the original gauge transformation of the theory, Equations \ref{trans1}-\ref{trans4}.  However, we can restore gauge invariance to the theory by introducing two phase fields, $\theta$ and $\phi$, which transform as:
\begin{subequations}
\begin{equation}
\theta\rightarrow \theta + \alpha,
\end{equation}
\begin{equation}
\phi\rightarrow\phi + \beta.
\end{equation}
\end{subequations}
These new fields couple to the gauge theory as:
\begin{align}
S =\int_{x,t}\bigg[\frac{1}{2} \hat{C}_{ijk\ell} E^{ij} E^{k\ell} - \frac{1}{2}\overline{\rho}^{-1}B^2 + \frac{1}{2} \overline{K}^{-1}e^2 -\frac{1}{2}\overline{\chi}^{-1} b^2 \nonumber \\
- \overline{g}B^ie_i - \underline{g}E^i_{\,i}b + \frac{1}{2}(c_1 (\partial_t\theta - A_0)^2 +c_2 (\partial_i\partial_j\theta - A_{ij})^2 \nonumber \\
+ c_3 (\partial_t\phi - a_0)^2 + c_4 (\partial_i\phi - a_i)^2) \bigg],
\end{align}
which is now a gauge-invariant action.  The previous form of the action can be obtained by gauge-fixing $\theta$ and $\phi$ to zero.  Note that, more properly, if we accounted for the discreteness of the charges and currents, the final four terms of the action above would all have a cosine form, $e.g.$ $\frac{1}{2}c_1(\partial_t\theta - A_0)^2\rightarrow - c_1\cos(\partial_t\phi - A_0)$, which is important within the uncondensed phase of the topological defects.

This completes the gauge dual description of a supersolid, from which several other dualities descend.  First, however, we turn to a more detailed analysis of the cross terms of our gauge dual, which have important consequences not only for the supersolid, but for the entire quantum phase diagram of bosons.

\subsection{Generalized Witten Effect and Symmetry Protected Subdimensionality}

In the gauge dual action of Equation \ref{supfindual}, we have ordinary ``$E^2 - B^2$" Maxwell terms for both the vector and tensor gauge fields, capturing the separate physics of particle-vortex and fracton-elasticity dualities.  Importantly, however, the action also contains ``cross terms" connecting the electric and magnetic fields of the two sectors.  Physically, these cross terms arise from the nontrivial coupling between the crystalline and superfluid sectors found in the original supersolid action ($i.e.$ the $g_1$ and $g_2$ terms of Equation \ref{St}).  We now ask what physical effects in a supersolid arise due to the presence of this coupling in the action.

To understand the role played by the cross terms, it is important to note that they bare a close similarity to the the $\vec{E}\cdot  \vec{B}$ term seen in axion electrodynamics.\cite{wilczek}  As such, we will refer to these extra couplings as generalized axion terms.  We have already shown that, as in conventional axion electrodynamics, the cross terms do not affect the physics of the gapless ($i.e.$ charge-free) sector of the theory.  Specifically, in the absence of topological defects, we found that the $g$ terms of the supersolid action do not enter the equations of motion.  However, these terms will have a significant effect on the charge sector of the theory.  To determine how the charge sector is altered, it is useful to recall the case of a conventional axion term, which produces a ``Witten effect," attaching electric charge to the magnetic monopoles of the theory.\cite{witten}  We expect similar physics to hold in the present case, except that the generalized axion terms should effect some charge attachment between the two sectors of the theory.

\begin{figure}[t!]
 \centering
 \includegraphics[scale=0.4]{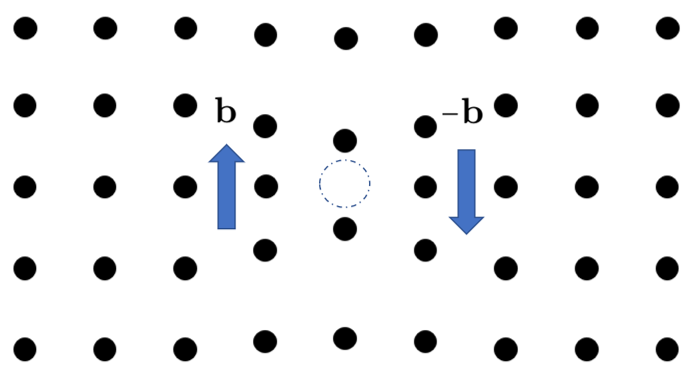}
 \caption{A bound state of two dislocations of opposite charge, separated by a single lattice constant, carries a unit of vacancy number, as can be seen by the depleted density of atoms.}
 \label{fig:vac}
 \end{figure}
 
To see the generalized Witten effect explicitly, we examine the Gauss's laws of the theory.  By varying the action with respect to $a_0$, we obtain the Gauss's law for the vector gauge field as:
\begin{equation}
\partial_ie^i = n_v - \overline{g}\partial_iB^i,
\end{equation} 
which indicates attachment of some magnetic flux of the tensor sector to the charges of the vector sector.  (Note that $\partial_iB^i$ of the noncompact tensor gauge field does \emph{not} correspond to a magnetic monopole configuration.)  In supersolid language, we have $\partial_iB^i = \epsilon^{ij}\partial_i\pi_j$, corresponding to the angular momentum associated with lattice displacements.  As such, this Gauss's law tells us that vortices of the vacancy/interstitial carry crystalline angular momentum.

It is even more informative to consider the Gauss's law of the tensor sector, obtained by varying the action with respect to $A_0$, yielding:
\begin{equation}
\partial_i\partial_jE^{ij} = s + \underline{g} \hat{C}^{-1}_{iik\ell}\partial^k\partial^\ell b,
\end{equation}
which represents a form of attachment of flux of the vector sector to charges of the tensor sector.  Recall that $b$ represents flux density of the vector gauge field, corresponding to density of vacancies/interstitials, not the magnetic field of the tensor gauge theory.  The presence of derivatives in the final term somewhat complicates the usual flux attachment interpretation.  The physics of this flux attachment is easiest to see on a lattice of high symmetry, such that $\hat{C}^{-1}_{iik\ell}\partial^k\partial^\ell b \sim \partial^2b$.  Given this diagonal second derivative structure, it is easy to verify that vacancies/interstitials are attached to \emph{quadrupoles} of the fracton charges, specifically quadrupoles corresponding to two head-to-head dipoles.  In elasticity language, this corresponds to the bound state of two dislocations, as seen in Figure \ref{fig:vac}.  This type of generalized Witten effect will have important consequences for the quantum phase diagram of bosons, as we will discuss later.
 
 \begin{figure}[t!]
 \centering
 \includegraphics[scale=0.35]{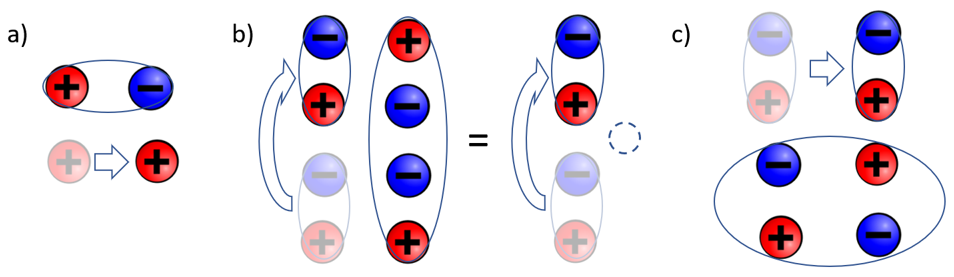}
 \caption{a) A fracton is immobile since motion of a fracton requires creation of a conserved dipole moment.  b) A dipole is immobile in the longitudinal direction in a phase without superfluid order, since such motion corresponds to creation of a collinear quadrupole, carrying conserved boson number, protected by global $U(1)$ symmetry.  c) A dipole is always fully mobile in the transverse direction, since it corresponds to creation of a $U(1)$-neutral non-collinear quadrupole moment.}
 \label{fig:quad}
 \end{figure}
 
This charge attachment physics is also important for understanding the mobility of lattice defects across the solid-supersolid transition.  In a generic tensor gauge theory without additional structure arising from elasticity duality, the quadrupoles are not conserved and thus dipoles are fully mobile.  Because of the generalized axion physics, however, quadrupoles with collinear charges carry a unit of boson number.  In phases with superfluid order, bosons can be freely created from the condensate, so all quadrupoles remain unconserved.  When boson number is conserved, however, such as in an ordinary commensurate crystal, collinear quadrupoles carry $U(1)$ atom charge, and are thus forbidden to be freely created from the vacuum.  Since longitudinal motion of dipoles corresponds to creation of collinear quadrupole moments (see Figure \ref{fig:quad}), the axion term forces dipoles to move only transversely when boson number is conserved.  This corresponds to the familiar glide constraint in commensurate crystals, which states that a dislocation can only move in the direction of its Burgers vector.  We therefore see that the one-dimensional constrained dynamics of dislocations only appears in the presence of the $U(1)$ symmetry associated with boson number conservation.  In contrast, a dislocation is fully mobile in a supersolid, in which this $U(1)$ symmetry is broken.  We therefore refer to this type of mobility restriction as symmetry-protected subdimensionality, or symmetry-enriched fracton order.  This example teaches us that the mobility of excitations in a fracton theory can be further reduced by the presence of global conservation laws, as has been noted in several related works.\cite{cheng,subsystem,potter}

\subsection{Mapping the Quantum Phase Diagram of Bosons}

 \begin{figure}[t!]
 \centering
 \includegraphics[scale=0.34]{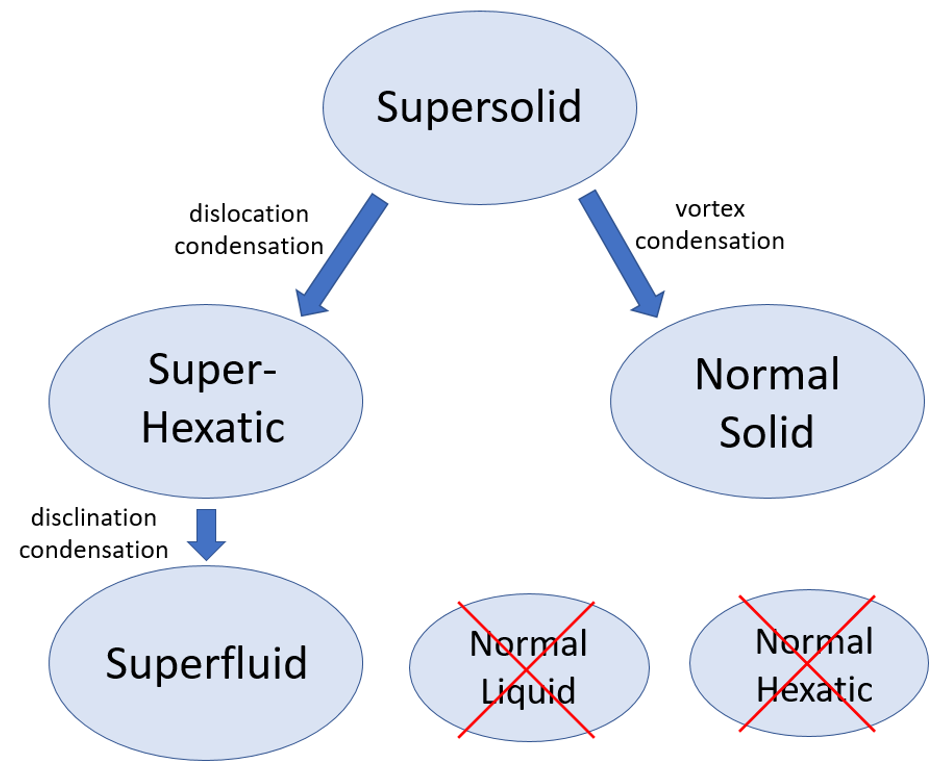}
 \caption{A supersolid features both superfluid order, with its associated confined vortex defects, and crystalline order, with its associated topological lattice defects, namely disclinations and dislocations.  Other typical quantum phases of bosons, such as normal solids, superfluids, and superhexatics, can be obtained from the supersolid upon condensation of lattice defects.  Note, however, that normal ($i.e.$ non-super) liquids and hexatics are ruled out as quantum phases on general principles from the LSM theorem.\cite{lsm,hastings,oshikawa}}
 \label{fig:descend}
 \end{figure}
 
\begin{figure}[t!]
 \centering
 \includegraphics[scale=0.38]{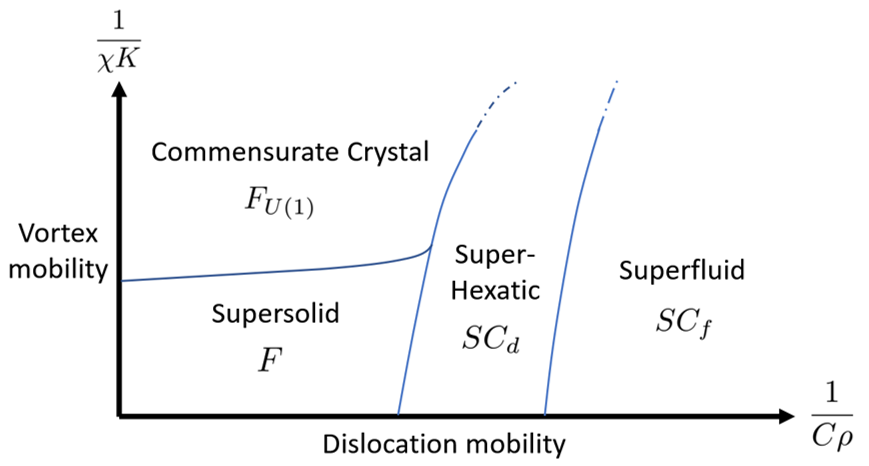}
 \caption{A schematic quantum ($i.e.$ zero-temperature) phase diagram of bosons in two dimensions.}
 \label{fig:phase}
 \end{figure}

We have now derived a gauge dual of a supersolid, which features both crystalline and superfluid order, as well as defects of both types of ordering, in the form of superfluid vortices and topological lattice defects.  By condensing these defects, we can destroy the different types of order and restore various symmetries to the system.  For example, by condensing the vortices, we can eliminate superfluid order and restore the underlying $U(1)$ symmetry of the system, leading to an ordinary commensurate crystal.  If, instead of vortices, we condense the topological lattice defects, we can also destroy crystalline order and restore spatial symmetries.  Specifically, by condensing dislocation defects, we restore translational order and enter a hexatic phase, in which rotational invariance is the only broken spatial symmetry.  By further condensing the disclination defects, we restore rotational invariance and enter into a liquid phase, with no broken spatial symmetries.  This sequence of condensation transitions is indicated in Figure \ref{fig:descend}.  We also sketch a schematic phase diagram of two-dimensional quantum systems of bosons in Figure \ref{fig:phase}.  Note that the hexatic and liquid phases obtained from the supersolid by condensing lattice defects continue to feature superfluid order, which is not affected by this condensation transition.

Indeed, superfluid order \emph{must} be present in any quantum liquid or hexatic phase, on general grounds, regardless of whether or not they are obtained from a supersolid.  These states descend from either a commensurate (normal) or incommensurate (supersolid) crystalline phase, which has been quantum melted by the condensation of topological lattice defects.  As we have just seen, such defects naturally carry boson number, even when the crystal does not possess superfluid order.  As such, the condensation of topological lattice defects naturally leads to condensation of bosons and the formation of  superfluid order.  This indicates that ``normal" ($i.e.$ non-superfluid) liquid and hexatic phases are impossible at zero temperature, which is consistent with conventional wisdom.  Indeed, the absence of a continuum quantum liquid phase of bosons without gapless modes or symmetry breaking can be argued based on the Lieb-Schultz-Mattis theorem\cite{lsm,hastings,oshikawa}, since a continuum system can simply be regarded as a lattice model in the limit where the lattice constant goes to zero.  As the continuum limit is approached, a lattice system necessarily passes through an infinite number of fillings, many of which require the system to be nontrivial ($i.e.$ symmetry-breaking, gapless, or topologically ordered).  This precludes the possibility of a stable trivial gapped phase in the continuum, since there are nontrivial phases arbitrarily close to the continuum limit.

It is worth noting that the structure of this theory is highly reminiscent of the physics of deconfined quantum criticality, in which there is a generic direct transition between two different types of symmetry breaking, in violation of the principles of Landau theory.\cite{dqcp,defined}  The most studied example is a direct transition from a N\'eel antiferromagnet to a valence bond solid, a continuous transition which has a critical point featuring an emergent gauge theory with deconfined excitations.  In theories such as this, the defects of one type of order carry the quantum numbers of a different type of order.  When one order parameter vanishes due to the condensation of its topological defects, another order parameter naturally arises due to the condensation of some appropriate quantum number.  Based on our gauge dual description of a supersolid, we see that a similar sort of structure occurs in ordinary two-dimensional boson systems.  If the commensurate crystal to superhexatic transition is continuous, it holds the possibility of hosting a deconfined quantum critical point.  Another possibility is the generic existence of a region of supersolid in the phase diagram between the commensurate crystal and superhexatic phases.  If the transition is indeed continuous and hosts a deconfined quantum critical point, then various other questions arise.  For example, can the field theory description of known deconfined quantum critical points, like the N\'eel-VBS transition, be phrased in a similar generalized axion language?  We leave these questions as topics for future investigation.

\section{Extensions}

In the previous sections, we have established dualities between fracton tensor gauge theories and certain familiar two-dimensional crystalline phases, such as ordinary commensurate crystals and supersolids.  However, this duality prescription is much more broadly applicable to quantum crystalline phases.  We here consider various extensions of our previous analysis to other phases breaking spatial symmetries.

\subsection{$T$-Breaking Crystals}

While the previously considered examples all featured time-reversal symmetry $T$, it is also possible to consider crystalline phases in contexts with explicitly broken time-reversal symmetry.  As simple examples, we can consider Abrikosov vortices in a type-II superconductor in a magnetic field, a neutral superfluid film under rotation, or a Wigner crystal of electrons in a magnetic field.  Lack of time-reversal invariance in such systems thus allows a new quadratic term in the low-energy quadratic theory of the form $\epsilon^{ij}u_i\partial_t u_j$.  Neglecting the less-relevant kinetic energy gives the low energy action for the chiral vortex crystal:
\begin{equation}
S = \int d^2xdt\frac{1}{2}\bigg(\epsilon^{ij}u_i\partial_tu_j - C^{ijk\ell}u_{ij}u_{k\ell}\bigg).
\end{equation}
Crucially, the new dynamic term $u_x\partial_t u_y$ encodes the appropriate cyclotron  vortex dynamics with $u_x$ and $u_y$ canonically conjugate, as is also found in electronic systems in a quantum Hall regime.\cite{tka}  As such, this action describes only a single phonon mode, with quadratic dispersion, $\omega\sim k^2$.  We now extend our previous analysis to find the gauge dual of such a chiral crystal, which was also derived independently in a parallel work by Kumar and Potter, using an imaginary time formalism.\cite{potter}

Once again, we begin by introducing Hubbard-Stratonovich fields $\pi_i$ and $\sigma_{ij}$, representing the lattice momentum and stress tensor, respectively.  As compared with the previous non-chiral treatment, these fields now enter the action in a more nontrivial way:
\begin{align}
S = \int d^2xdt\bigg(\frac{1}{2}C_{ijk\ell}^{-1}\sigma^{ij}\sigma^{k\ell}& - \frac{1}{2}\epsilon^{ij}\pi_i\partial_t\pi_j \nonumber\\
&- \sigma^{ij}u_{ij} + \pi_i\partial_tu_i\bigg),
\end{align}
where the original action is obtained upon integrating out the new fields.  As before, we now break up the displacement field into its smooth $\tilde{u}_i$ and singular $u_i^{(s)}$ components, where $\tilde{u}_i$ is single-valued and $u_i^{(s)}$ serves as a source for topological defects.  As in a conventional crystal, integration of the phonons enforces the Newton equation:
\begin{equation}
\partial_t\pi^i - \partial_j\sigma^{ij} = 0,
\end{equation}
that with $B^i = \epsilon^{ij}\pi_j$ and $E_\sigma^{ij} = \epsilon^{ik}\epsilon^{j\ell}\sigma_{k\ell}$ again maps onto the Faraday's law:
\begin{equation}
\partial_t B^i + \epsilon_{jk}\partial^jE^{ki}_\sigma = 0,
\end{equation}
which is solved by $B^i = \epsilon_{jk}\partial^jA^{ki}$ and $E^{ij}_{\sigma} = -\partial_tA^{ij} - \partial^i\partial^j\phi$.  In terms of these fields, the action can now be written in dual form as:
\begin{equation}
\int d^2xdt\bigg(\frac{1}{2}\tilde{C}^{-1}_{ijk\ell}E^{ij}_\sigma E^{k\ell}_\sigma - \frac{1}{2}\epsilon^{ij}B_i\partial_tB_j + \rho\phi - J^{ij}A_{ij}\bigg),
\end{equation}
where the fracton charge density $\rho$ and current $J^{ij}$ are defined in the same way as before.  The only notable change in this dual gauge theory, as compared with the non-chiral crystal, is the absence of the usual $B^2$ term in favor of a $B\partial_tB$ contribution to the action.  This change results in only a single gapless gauge mode with a quadratic dispersion, $\omega\sim k^2$, which matches with the properties of the chiral crystal.

\subsection{Superhexatic}

We previously discussed the gauge dual of a supersolid, which features both a tensor gauge field describing the crystalline sector and a vector gauge field describing the superfluid sector.  Upon condensing vortices of the superfluid, we obtain the pure tensor gauge theory of fracton-elasticity duality.  And upon condensing all topological lattice defects, both disclinations and dislocations, we obtain the familiar vector gauge theory of particle-vortex duality.  However, condensing only dislocations (while leaving disclinations uncondensed and gapped), leads to a zero-temperature quantum hexatic phase, the thermal analogue of which was discovered by Halperin and Nelson.\cite{halperin}  Specifically, as discussed earlier, such a condensation at zero temperature will necessarily lead to a superhexatic phase featuring both hexatic and superfluid order.  But what sort of gauge theory is dual to such a superhexatic phase?

One approach to answering this question is from the dual formulation of a supersolid, featuring both vector and tensor gauge theories, then explicitly condensing dislocation defects.  However, a simpler path is to identify the relevant low-energy modes and write down the most general quadratic field theory.  For a two-dimensional superhexatic, the important degrees of freedom are the Goldstone modes of the associated orientational and U(1) atom conservation symmetries, which we denote as $\theta$ and $\phi$, respectively.  The action should be invariant under shifts of $\theta$ and $\phi$ by global constants, so only derivatives of these fields should appear in the action.  It is also important that $\phi$ is even under spatial reflections and odd under time reversal, while $\theta$ has the opposite behavior.  Making use of these facts, the most general action which can be written down for a superhexatic, to quadratic order in the fields and derivatives, takes a simple decoupled form:
\begin{align}
S = \int d^2xdt\frac{1}{2}\bigg((\partial_t\phi)^2 - (\partial_i\phi)^2 + (\partial_t\theta)^2 - (\partial_i\theta)^2\bigg),
\end{align}
which is simply two XY models, as expected on general grounds, with coupling only appearing as anharmonic gradient interactions.\cite{superhex}  In contrast to the case of the supersolid, there are no symmetry-allowed cross terms coupling the $\theta$ and $\phi$ sectors at the quadratic level.  At this level, we can therefore trivially construct a gauge dual for the superhexatic by writing down separate vector gauge theories for the two sectors:
\begin{equation}
S = \int d^2xdt\frac{1}{2}\bigg(E^iE_i - B^2 + e^ie_i - b^2\bigg),
\end{equation}
where the electric and magnetic fields have their usual Maxwell definition in terms of two vector gauge fields, $A_i$ and $a_i$.  Let us take $A_i$ as the dual gauge field of the orientational sector and $a_i$ as the dual gauge field of the superfluid sector.  Then we can conclude that $e_i$ and $B$ are odd under time-reversal and spatial reflections, while $E_i$ and $b$ are even under these symmetries.

We have now shown that the gauge dual of a superhexatic takes an extremely simple form, featuring two conventional vector gauge theories which are completely decoupled at the quadratic level, coupling only via higher order terms.  This gauge dual hosts two gapless gauge modes with linear dispersion, corresponding to the Goldstone modes of the two types of symmetry breaking.

\subsection{Fermionic Atoms}

In Section \ref{sec:comp}, we incorporated the statistics of atoms into the crystalline duality, assuming the atoms of the crystal were bosonic.  A natural extension is to a crystal of fermionic atoms.  This opens the door to a variety of new quantum phases with no bosonic analogue.  An obvious example is a state in which the fermionic vacancy/interstitial defects have formed a Fermi surface.  Just as the supersolid served as a parent state for the entire bosonic quantum phase diagram, the ``Fermi surface solid" serves as a parent state for all phases of fermionic atoms.

To write down an effective theory of the Fermi surface crystal, it is useful to break the Fermi surface up into patches $n$ over which the normal direction $\perp$ does not change substantially.  Within this framework, one can write the low-energy action for a noninteracting system of fermions as\cite{nodal}:
\begin{equation}
S_\Psi = \sum_{\textrm{patches}}\int dtd^2x\,\overline{\Psi}_n(\partial_t - v_F \sigma_x\partial_{x_\perp})\Psi_n,
\label{psiact}
\end{equation}
where $\Psi_n$ is a Weyl fermion for each patch of the Fermi surface, $x_\perp$ is the normal coordinate in real space at each patch, and $\sigma_x$ acts on the spin space of the fermions.  This model could also be extended to include interactions between different patches to yield the celebrated Landau's Fermi liquid theory.  For now, however, we content ourselves with minimally coupling this Fermi surface to the crystalline degrees of freedom.  As in the case of the supersolid, the symmetry-allowed couplings between the two sectors take the form of density-density and current-current interactions, which we can write as:
\begin{equation}
S_{u\Psi} = g\sum_{\textrm{patches}}\int dtd^2x\,(\partial_iu^i\overline{\Psi}_n\Psi_n - \partial_tu^i\overline{\Psi}_n\partial_i\Psi_n),
\label{upsiact}
\end{equation}
where the density-density and current-current interactions must appear with opposite signs, as discussed in the context of the supersolid.  A similar coupling between crystalline degrees of freedom and a Fermi surface was also discussed in Reference \onlinecite{potter} in the context of charge density waves.  The full field theory of the Fermi surface solid can then be written down as:
\begin{equation}
S = S_\Psi + S_u + S_{u\Psi},
\end{equation}
where $S_u$ is the usual elastic action, and $S_\Psi$ and $S_{u\Psi}$ are defined in Equations \ref{psiact} and \ref{upsiact}, respectively.

As in the case of a supersolid, it is useful to formulate this theory in gauge dual language.  However, constructing gauge duals of fermionic phases is a highly nontrivial task, an understanding of which has only begun to emerge in the last several years, beginning in the context of surface states of topological insulators.\cite{fermidual1,fermidual2,web}  Instead of trying to dualize the full theory, we therefore content ourselves with dualizing only the crystalline sector, leaving the Fermi surface sector of the theory in its original language.  From our earlier analysis, we can easily construct this dual theory in the usual way, introducing a tensor gauge field to describe the dynamics of the crystalline sector.  The resulting theory is described by the following action:
\begin{align}
&S = \int d^2xdt\,\frac{1}{2}(\tilde{C}^{-1}_{ijk\ell}E_\sigma^{ij}E_\sigma^{k\ell} - B^iB_i) + S_\Psi \nonumber\\
&+g\sum_{\textrm{patches}}\int dtd^2x\,(E^{ii}_\sigma\overline{\Psi}_n\Psi_n - B^i\overline{\Psi}_n\partial_i\Psi_n) + \cdot\cdot\cdot,
\end{align}
where $E_\sigma^{ij} = -\partial_tA^{ij} - \partial^i\partial^j\phi$ and $B^i = \epsilon_{jk}\partial^jA^{ki}$, as before, and ``$\cdot\cdot\cdot$" represents source terms, which we have suppressed.  By varying the action with respect to $\phi$, the Gauss's law of the theory is given by:
\begin{equation}
\partial_i\partial_jE^{ij} = \rho + g\partial^2\sum_{\textrm{patches}}\overline{\Psi}_n\Psi_n.
\end{equation}
As with bosonic atoms, the second derivative above indicates that fermion number is attached to linear quadrupoles of the disclinations (fractons).

This type of charge attachment imposes that a dislocation of the crystalline order can only move in the direction of its Burgers vector ($i.e.$ perpendicular to their dipole moment in the gauge theory language), while motion perpendicular to the Burgers vector results in creation of a fermionic vacancy/interstitial defect.  More mathematically, a bare dislocation hopping operator, $\vec{b}^\dagger_{x+p}\vec{b}_x$ (where $p$ is the perpendicular direction to $b$), is not allowed.  Instead, an allowed hopping operator takes the form:
\begin{equation}
\vec{b}^\dagger_{x+p}\vec{b}_x\Psi^\dagger_x,
\end{equation}
featuring a combination of dislocation hopping and fermion creation.  Interestingly, since this operator must preserve the fermion parity of the system, we see that the hopping operator $\vec{b}^\dagger_{x+p}\vec{b}_x$ must be a \emph{fermionic} operator, in contrast to the bosonic statistics of all conventional hopping operators, as first noted in Reference \onlinecite{potter}.  While an unusual property, the fermionic nature of the hopping operator does not appear to be a fundamental impediment to the formulation of the theory.  Indeed, we conjecture that such statistics associated with hopping operators may prove useful in the more general classification of the statistics of fractons and subdimensional particles.

We have already seen how the supersolid serves as a parent state for all conventional bosonic phases of matter, yielding the ordinary commensurate crystal and superfluid states upon condensation of various defects.  Similarly, we expect to be able to access various phases of fermionic atoms starting from the Fermi surface solid.  Melting the crystalline order will eventually result in a Landau Fermi liquid, with perhaps a Fermi surface hexatic as an intermediate state.  By Cooper pair condensation, we can also obtain phases featuring superconductivity, which can coexist with either hexatic or full crystalline order.  As with bosons, the structure of the duality does not permit a trivial gapped phase preserving all symmetries.  We leave the full mapping of the fermionic quantum phase diagram as a challenge for future work.

\subsection{Other Extensions}

Before moving on, we also briefly describe several other extensions to our work which have appeared in subsequent literature.  Perhaps most notably, the first duality between fracton tensor gauge theory and ordinary two-dimensional quantum crystals has now been extended to three dimensions.\cite{pai}  Three-dimensional crystals have the nontrivial feature that their topological lattice defects, such as disclinations and dislocations, are line-like objects, rather than the point-like defects found in two-dimensional crystals.  Accordingly, these systems require a more complicated rank-4 gauge dual formulation, which turns out to combine the properties of symmetric tensor gauge fields with those of \emph{anti}symmetric tensor gauge fields, the latter of which naturally hosts line-like excitations.  In this way, the gauge dual of three-dimensional crystals describes line-like topological defects which inherit some of the mobility restrictions of fractons.  Specifically, these line-like objects obey higher-moment conservation laws on their flux through arbitrary two-dimensional surfaces, in close analogy with higher-form (or ``generalized") symmetries.  These conservation laws force the disclination lines of a three-dimensional crystal to be fully immobile in isolation, leading us to label this type of excitation as a fractonic line.

In another development, it has also been argued that the elasticity theory of certain chiral systems may be described by a gapped tensor gauge theory, taking the form of a higher-rank Chern-Simons theory.\cite{gromov}  Such theories may have applicability to Chern insulators or topological metamaterials, within certain limits.  These theories are notable for exhibiting fractionalization of the ordinary topological lattice defects, such as fractional Burgers vectors.  This proposal may open the door to a new realm of topological elasticity theories, which are an exciting topic of future investigation.  In turn, such topological elasticity theories may provide clues in the search for two-dimensional lattice models realizing fracton physics, which remain elusive.

\section{Applications}

With this new duality in hand, relating elasticity theory to fracton tensor gauge theories, we explore a few applications of this duality.  In this section, we show how fracton-elasticity duality can be utilized both to make new theoretical predictions about fracton systems, and also to provide a simpler derivation of the known properties of two-dimensional melting transitions.

\subsection{Two-Dimensional Thermal Melting}

The previous sections have shown how to reformulate the conventional theory of elasticity as a dual gauge theory.  As an important check of this framework, we can use it to reproduce and simplify the description of classical two-stage melting of a two-dimensional crystal, through a thermal hexatic to an isotropic liquid, as first discussed by Halperin and Nelson\cite{halperin,nelson} and by Young.\cite{young}  To this end, we first rederive the duality in a classical context, which provides certain simplifications over the full quantum case.  This leads to a generalized vector sine-Gordon model, on which a renormalization group analysis was performed in Reference \onlinecite{z3} to obtain the critical exponents of the solid-to-hexatic melting transition, driven by the proliferation of dislocation defects.  Here we discuss some of the basis aspects of this analysis, which complements the traditional vector Coulomb gas treatment of Halperin, Nelson, and Young.\cite{halperin,nelson,young}

\subsubsection{Classical Duality}

The starting point for the classical version of our duality is a classical Hamiltonian featuring only the linearized potential energy associated with configurations of the lattice displacement, $H = \int d^2x\mathcal{H}[{\bf u}]$, with Hamiltonian density given by:
\begin{equation}
\mathcal{H}[{\bf u}] = \frac{1}{2}C^{ijk\ell}u_{ij}u_{k\ell},
\end{equation}
where $u_{ij} = \frac{1}{2}(\partial_iu_j + \partial_ju_i)$ is the usual symmetric strain tensor.  (Note that, at the classical linearized level, the kinetic piece, $\frac{1}{2}\pi^2$, can be integrated out and has no bearing on the subsequent analysis.)  The form of the elastic tensor $C^{ijk\ell}$, including its number of independent components, is dictated by the symmetry of the underlying crystal.  For simplicity, we here focus on the case of an isotropic hexagonal lattice, for which this tensor takes the generic form:
\begin{equation}
C^{ijk\ell} = \lambda \delta^{ij}\delta^{k\ell} + 2 \mu\delta^{ik}\delta^{j\ell},
\end{equation}
characterized by two independent elastic constants $\lambda$ and $\mu$, known as the Lam\'e coefficients.  Just as in the full quantum case, the field $u_i$ includes a smooth phonon piece and a singular piece from topological defects, the latter of which encodes disclinations according to:
\begin{equation}
\epsilon^{i\ell}\epsilon^{jk}\partial_\ell\partial_ku_{ij} = \frac{2\pi}{n}s_{tot}\equiv\rho,
\end{equation}
where the total disclination charge is $\frac{2\pi}{n}s_{tot} = \frac{2\pi}{n}s + \epsilon^{ij}\partial_ib_j$, where $s$ is the density of bare disclinations and the second term is the contribution from the dislocation density $b_i$.  This definition of a disclination leads to the bond angle $\theta = \frac{1}{2}\epsilon^{ij}\partial_iu_j$ winding by $2\pi/n$ upon going around the defect, as discussed earlier in the quantum case.  Once again, there are also stable dipolar bound states of disclinations, corresponding to dislocations (see Figure \ref{fig:disl}), which are defects of the translational order.

We consider the thermal partition function:
\begin{equation}
Z = \int \mathcal{D}{\bf u}\,e^{-\int d^2x\mathcal{H}[{\bf u}]},
\end{equation}
where we have set $\beta = (k_BT)^{-1} = 1$ for simplicity ($i.e.$ we measure all coupling constants in units of temperature).  We now introduce a Hubbard-Stratonovich field $\sigma_{ij}$, which physically plays the role of the stress tensor, to write the partition function as:
\begin{equation}
Z = \int\mathcal{D}{\bf u}\mathcal{D}\sigma_{ij}\,e^{-\int d^2x\mathcal{H}[{\bf u},\sigma_{ij}]},
\end{equation}
where the new Hamiltonian density is given by:
\begin{align}
\mathcal{H}[{\bf u},\sigma_{ij}] &= \frac{1}{2}C^{-1}_{ijk\ell}\sigma^{ij}\sigma^{k\ell} + i\sigma^{ij}u_{ij}\nonumber\\
&= \frac{1}{2}C^{-1}_{ijk\ell}\sigma^{ij}\sigma^{k\ell} + i\sigma^{ij}(\partial_i\tilde{u}_j + u_{ij}^{(s)}),
\end{align}
where we have broken up the symmetric strain tensor into its contributions from the smooth phonon field $\tilde{u}_i$ and from topological defects $u_{ij}^{(s)}$.  The original partition function is obtained upon integrating out the new field $\sigma_{ij}$.  Since the smooth single-valued piece $\tilde{u}_i$ appears linearly in the Hamiltonian, with imaginary coefficient, it can be integrated out of the problem to enforce a divergenceless constraint on the stress tensor:
\begin{equation}
\partial_i\sigma^{ij} = 0.
\label{divcond}
\end{equation}
This constraint can be solved explicitly by a scalar potential $\phi$ representing the Airy stress function:
\begin{equation}
\sigma_{ij} = \epsilon_{ik}\epsilon_{j\ell}\partial^k\partial^\ell\phi.
\end{equation}
Note that, in terms of our previous quantum duality, this would correspond to an electrostatic potential formulation:
\begin{equation}
E^{ij}_\sigma = \partial^i\partial^j\phi,
\end{equation}
which makes intuitive sense, since Equation \ref{divcond} is simply the static limit of Faraday's equation.  We now express the Hamiltonian density in terms of this potential function, obtaining:
\begin{equation}
\mathcal{H}[\phi] = \frac{1}{2}\tilde{C}^{-1}_{ijk\ell}\partial^i\partial^j\phi\partial^k\partial^\ell\phi + i\epsilon^{ik}\epsilon^{j\ell}\partial_k\partial_\ell\phi u_{ij}^{(s)},
\end{equation}
where we have defined $\tilde{C}_{ijk\ell} = \epsilon_{ia}\epsilon_{jb}\epsilon_{kc}\epsilon_{\ell d}C^{abcd}$.  Integrating by parts on the second term, and using the definition of disclination density, we can rewrite the Hamiltonian as:
\begin{align}
\mathcal{H}[\phi] = \frac{1}{2}\tilde{C}^{-1}_{ijk\ell}\partial^i\partial^j\phi\partial^k\partial^\ell\phi + i\phi\bigg(\frac{2\pi}{6}s + \epsilon^{ij}\partial_ib_j\bigg).
\end{align}
where we have specialized to a hexagonal crystal, with $n=6$.

Within the crystal phase, the disclination defects are extremely energetically costly and have little relevance to transitions out of the phase.  Neglecting these defects for now ($i.e.$ setting $s=0$), we can straightforwardly integrate the field $\phi$ out of the partition function to obtain the energy as a function of the dislocation configurations:
\begin{equation}
H_{\bf b} = \frac{1}{2}\int\frac{d^2q}{(2\pi)^2}b^i({\bf q})\tilde{K}_{ij}({\bf q})b^j(-{\bf q}),
\end{equation}
with the tensor interaction $\tilde{K}_{ij}$ given in momentum and coordinate spaces by:
\begin{subequations}
\begin{eqnarray}
\tilde{K}_{ij}({\bf q}) &=& \frac{K}{q^2}\bigg(\delta_{ij} - \frac{q_iq_j}{q^2}\bigg),\\
 K_{ij}({\bf r})&=&-\frac{K}{4\pi} 
\left(\delta_{ij}\ln(r/a) -\frac{r_i r_j}{r^2}\right),
\end{eqnarray}
\end{subequations}
where we have defined the elastic modulus $K = \frac{4\mu(\mu+\lambda)}{2\mu + \lambda}$.  Converting back to real space, the effective Hamiltonian for the dislocations corresponds to that of a vector Coulomb gas:
\begin{align}
H_{\bf b} = &-\frac{K}{8\pi}\int d^2xd^2y\bigg(b^i({\bf x})b_i({\bf y})\ln\frac{|{\bf x}-{\bf y}|}{a}\nonumber\\
&- \frac{b^i({\bf x})({\bf x}-{\bf y})_ib^j({\bf y})({\bf x}-{\bf y})_j}{|{\bf x}-{\bf y}|^2}\bigg),
\label{bham}
\end{align}
where $a$ is the lattice constant. Precisely this same vector Coulomb gas was used by Halperin and Nelson\cite{halperin,nelson} and by Young\cite{young} in their seminal work on two-dimensional melting, demonstrating the equivalence of the dual framework with their analysis.

Instead of integrating out $\phi$ to obtain the effective Hamiltonian for topological defects, we can take the complementary but equivalent approach of summing over defect configurations to obtain an effective Hamiltonian for $\phi$.  To this end, we explicitly write the disclination and dislocation densities in terms of their discrete charges:
\begin{equation}
b^i({\bf x}) = \sum_{{\bf x}_n} b^i_{{\bf x}_n}\delta^{(2)}({\bf x}-{\bf x}_n),
\end{equation}
\begin{equation}
s({\bf x}) = \sum_{{\bf x}_n} s_{{\bf x}_n} \delta^{(2)}({\bf x}-{\bf x}_n),
\end{equation}
where ${\bf x}_n=a(n_1 \hat{\bf e}_1+n_2 \hat{\bf e}_2)$, ($n_1, n_2
\in \mathbb{Z}$) are triangular lattice vectors spanned by unit
vectors $\hat{\bf e}_1= \hat{\bf x}$ and $\hat{\bf
  e}_2=\frac{1}{2}\hat{\bf x}+\frac{\sqrt{3}}{2} \hat{\bf y}$, ${\bf
  b}_{{\bf x}_n}=a(n_1 \hat{\bf e}_1+n_2 \hat{\bf e}_2)$, and $s_{{\bf
    r}_n} \in \mathbb{Z}$ are dislocation and disclination charges,
respectively.

Expressing the Hamiltonian of Eq. \ref{bham} in terms of these charges, we find:
\begin{align}
H = \frac{1}{2}\int d^2x\tilde{C}^{-1}_{ijk\ell}\partial^i\partial^j\phi\partial^k\partial^\ell\phi + \sum_{{\bf x}_n}\bigg[\tilde{E}_{b}{\bf b}^2_{{\bf x}_n} + E_{s}s_{{\bf x}_n}^2\bigg]\nonumber\\
-i\sum_{{\bf x}_n}\bigg[\epsilon_{ij}\partial^i\phi({\bf x}_n)b^j_{{\bf x}_n} - \frac{2\pi}{6}\phi({\bf x}_n)s_{{\bf x}_n}\bigg],
\end{align}
where we have added by hand core energies $E_{b}= a^2\tilde E_{b}$ and $E_{s}$ for the disclination and dislocation  respectively, to account for their short-distance energetics.  Summing over the fundamental ($i.e.$ charge $\pm 1$) topological defects $\{s_{{\bf x}_n},{\bf b}_{{\bf x}_n}\}$ inside the partition function gives, via standard analysis\cite{chaikin,z3}, $Z = \int\mathcal{D}\phi\,e^{-\tilde{H}}$ with the effective Hamiltonian:
\begin{subequations}
\begin{align}
\tilde{H} = \int d^2&x\bigg[\frac{1}{2}\tilde{C}^{-1}_{ijk\ell}\partial^i\partial^j\phi\partial^k\partial^\ell\phi\nonumber\\
&-g_b\sum_{n=1}^3\cos(\epsilon_{ij}b^i_n\partial^j\phi) - g_s\cos\bigg(\frac{2\pi}{6}\phi\bigg)\bigg],\\
= \int d^2&x\bigg[\frac{K^{-1}}{2}(\partial_i\partial_j\phi)^2 + \frac{B}{2}(\partial^2\phi)^2\nonumber\\
&-g_b\sum_{n=1}^3\cos(\epsilon_{ij}b^i_n\partial^j\phi) - g_s\cos\bigg(\frac{2\pi}{6}\phi\bigg)\bigg],
\label{vecsingord}
\end{align}
\end{subequations}
where $n$ runs over the three elementary dislocation Burgers vectors, given by ${\bf b}_1=a\hat{\bf x}, {\bf b}_2=-\frac{a}{2} \hat{\bf
  x}+\frac{a\sqrt{3}}{2}\hat{\bf y}, {\bf b}_3=-{\bf b}_1-{\bf
  b}_2=-\frac{a}{2} \hat{\bf x}-\frac{a\sqrt{3}}{2}\hat{\bf y}$.  The parameters $K^{-1}$ and $B$ can be written in terms of the Lam\'e coefficients as $K^{-1} = \frac{2\mu + \lambda}{4\mu(\mu + \lambda)}$ and $B = \frac{\lambda}{4\mu(\mu+\lambda)}$.  The coupling constants are given by $g_b = \frac{2}{a^2}e^{-E_{b} }$ and $g_2 = \frac{2}{a^2}e^{-E_{s}}$, corresponding to the fugacities of the dislocations and disclinations, respectively.  This generalized vector sine-Gordon model provides a complete characterization of a two-dimensional crystal and allows a complementary treatment of its two-stage melting.

\subsubsection{Renormalization Group Analysis}

As we demonstrated in the context of the quantum duality, dislocations have a logarithmic interaction energy, $\ln(R/a)$, where $R$ is the separation of two dislocations.  As in the conventional BKT transition\cite{ber1,ber2,kt}, this admits an entropy-driven unbinding of dislocations, and the accompanied vanishing of the shear modulus and restoration of the translational symmetry, $i.e.$ melting into a hexatic fluid.  In contrast, disclinations have a quadratic interaction energy, $R^2$, and therefore at this stage remain bound, unless the continuous transition is preempted by a first order melting.  This can be seen in a complementary way through power-counting on the $g_s$ operator at the Gaussian fixed line.  This shows that as long as $g_b$ is small, $g_s$ is strongly irrelevant with scaling dimension $-2$.\cite{z3}

To study this crystal to hexatic fluid melting transition, we can entirely neglect disclinations, which remain bound across the transition.  This allows us to simply drop the disclination piece of the vector sine-Gordon model, $i.e.$ set $g_s = 0$, with the effective Hamiltonian reducing to:
\begin{align}
\tilde{H} = \int d^2x\bigg[\frac{1}{2}&\tilde{C}^{-1}_{ijk\ell}\partial^i\partial^j\phi\partial^k\partial^\ell\phi\nonumber\\
&-g_b\sum_{n=1}^3\cos(\epsilon_{ij}b^i_n\partial^j\phi)\bigg],
\label{lapelast}
\end{align}
a vector sine-Gordon model that is the starting point for our renormalization group analysis.  To simplify notation, it is useful to define a new field given by:
\begin{equation}
\mathcal{A}^i = \epsilon^{ij}\partial_j\phi,
\end{equation}
which obeys the divergence-free condition, $\partial_i\mathcal{A}^i = 0$, with the model reducing to:
\begin{align}
H = \int d^2x&\bigg[\frac{1}{2}C^{-1}_{ijk\ell}\epsilon^{im}\epsilon^{kn}\partial_m\mathcal{A}^j\partial_n\mathcal{A}^\ell\nonumber\\
&+\frac{\alpha}{2}(\partial_i\mathcal{A}^i)^2 - g_b\sum_{n=1}^3\cos(b^i_n\mathcal{A}_i)\bigg],
\end{align}
where we have chosen to impose the divergence-free condition energetically via a $(\partial_i\mathcal{A}^i)^2$ term, with the coefficient $\alpha$ taken to infinity at the end of the calculation.  Specializing to the case of the hexagonal lattice, we can also expand the Hamiltonian as:
\begin{align}
H = \int d^2x\bigg[\frac{K^{-1}}{2}(\partial_i\mathcal{A}_j)^2 + \frac{B}{2}\partial_i\mathcal{A}_j\partial^j\mathcal{A}^i + \frac{\alpha}{2}(\partial_i\mathcal{A}^i)^2 \nonumber\\
-g_b \sum_{n=1}^3\cos(b^i_n\mathcal{A}_i)\bigg].
\end{align}
In the physical limit $\alpha\rightarrow\infty$, the $\mathcal{A}^i$ propagator in the dislocation-free sector takes the purely transverse form:
\begin{equation}
\langle \mathcal{A}_i({\bf q})\mathcal{A}_j({\bf q}')\rangle_0 = \frac{K}{q^2}(2\pi)^2\delta^{(2)}({\bf q}+{\bf q}')P^T_{ij}({\bf q}),
\end{equation}
where the transverse projection operator is given by $P^T_{ij}({\bf q}) = \delta_{ij} - \frac{q_iq_j}{q^2}$.  While a perturbative expansion in $g_b$ is convergent at low temperatures, deep within the crystal phase, a naive perturbation theory breaks down in the vicinity of the crystal-hexatic transition.  To understand the physics of the critical point, we must perform a renormalization group analysis.  Specifically, we carry out a Wilsonian momentum-shell RG treatment, breaking up $\mathcal{A}^i$ into its slow and fast modes as:
\begin{subequations}
\begin{equation}
\mathcal{A}_i^<({\bf x}) = \int_{0<q<\Lambda/b}\frac{d^2q}{(2\pi)^2}e^{i{\bf q}\cdot {\bf x}}\mathcal{A}_i({\bf q}),
\end{equation}
\begin{equation}
\mathcal{A}_i^>({\bf x}) = \int_{\Lambda/b<q<\Lambda}\frac{d^2q}{(2\pi)^2}e^{i{\bf q}\cdot {\bf x}}\mathcal{A}_i({\bf q}),
\end{equation}
\end{subequations}
where we have taken the UV cutoff as $\Lambda = 2\pi/a$ with $b>1$ as the coarse-graining factor.  Integrating the short-scale modes out of the partition function yields an effective Hamiltonian for the long-scale modes.  We refer the reader to Reference \onlinecite{z3} for details of the calculation.  Here we simply state the results that the renormalized couplings for the slow modes are given by:
\begin{subequations}
\begin{align}
K_R^{-1}(b) &= K^{-1} + J_2 g_b^2,\\
B_{R}(b) &= B + J_3 g_b^2,\\
g_{b,R}(b) &= g_be^{-\frac{1}{2}G^>_{nn}(0)} + J_1 g_b^2,
\end{align}
\end{subequations}
valid to second-order in $g_b$, where the $J_i$ are numerical factors defined in terms of modified Bessel functions, while the Green's function appearing in the definition of $g_{b,R}$ is given by $G_{nm}^>({\bf x}-{\bf y}) \equiv b^i_nb^j_m\langle \mathcal{A}_i^>({\bf x})\mathcal{A}_j^>({\bf y})\rangle^>_0$.  These RG equations can now be converted into differential equations by taking $b = e^{\delta\ell}$ with $\delta\ell\ll 1$.  In the vicinity of the fixed point at $g_b^*=B^* =0$, $K_R^{-1,*} = \frac{a^2}{16\pi}$, the differential RG flow equations for the dimensionless couplings $\overline{K}^{-1}(\ell)=\frac{K_R^{-1,*}(l)}{a^2}$, $\overline{B}(\ell)=\frac{B(l)}{a^2}$, and $\overline{g}_b(\ell) = g_b(\ell)a^2$ take the form:
\begin{subequations}
\begin{align}
\frac{d\overline{K}^{-1}(\ell)}{d\ell} &=\frac{3\pi}{8} \bigg[ e^2\bigg(I_0(2) - \frac{1}{2}I_1(2)\bigg)\bigg]\overline{g}_b^2(\ell),\\
\frac{d\overline{B}(\ell)}{d\ell} &= \frac{3\pi}{16}e^2I_1(2)\overline{g}_b^2(\ell),\\
\frac{d\overline{g}_b(\ell)}{d\ell} &= \left(2-\frac{\overline{K}}{8\pi}\right)\overline{g}_b+\pi e I_0(2)\overline{g}_b^2(\ell),
\end{align}
\end{subequations}
where $I_0(x)$ and $I_1(x)$ are modified Bessel functions.  Using the definitions of $\overline{K}^{-1}$, $\overline{B}$, and $\overline{g}_b$ in terms of  the dimensionless Lam\'e elastic
constants $\overline{\mu}=\mu a^2$, $\overline{\lambda}=\lambda a^2$
and the fugacity $y=e^{-E_b}$, we recover precisely RG flows for the inverse
shear modulus, $\overline{\mu}^{-1}(l)$, inverse bulk modulus $[\overline{\mu} (l)
+\overline{\lambda}(l)]^{-1}$, and the effective fugacity $y(l)$ respectively,
\begin{subequations}
\begin{eqnarray}
  \frac{d\overline{\mu}^{-1}}{dl}&=& 3\pi e^2 I_0(2) y^2, \\
  \frac{d (\overline{\mu}+\overline{\lambda})^{-1}} {dl}&=& 3\pi e^2 \left[I_0(2)-I_1(2)\right] y^2,\\
  \frac{d y} {dl}&=& \left(2-\frac{\overline{K}}{8\pi}\right)y +2\pi e I_0(2)y^2,
\end{eqnarray}
\end{subequations}
first derived by Halperin and Nelson\cite{halperin,nelson}, and by Young.\cite{young}

Following a standard analysis, as first shown by Halperin and Nelson, these RG equations can be used to obtain a characteristic correlation length $\xi_1$ near the crystal-hexatic critical temperature, $T_{c1}$, which is given by:
\begin{equation}
\xi_1(T) \sim ae^{-c/|T-T_{c1}|^{\nu}},
\label{thermlength}
\end{equation}
where the exponent $\nu$ is given by $0.3696...$ on a hexagonal lattice and $c$ is a non-universal constant.

We can also now examine the criticality at the hexatic-liquid transition.  In the hexatic phase, dislocations have condensed and therefore screen disclinations.  This is captured by the large relevant dislocation fugacity, $g_b\gg 1$, where the vector cosine operator from the original Hamiltonian (Equation \ref{vecsingord}) can be treated within the harmonic approximation, $i.e.$ $\cos(\epsilon_{ij}b^i_n\partial^j\phi)\rightarrow 1-\frac{1}{2}(\epsilon_{ij}b^i_n\partial^j\phi)^2$, resulting in a more standard gradient ``elasticity" for the Airy potential $\phi$.  In its presence, we can neglect the Laplacian elasticity in the Hamiltonian of Eq. \ref{lapelast}, which is subdominant at long scales, resulting in a conventional sine-Gordon model for $\phi$ with Hamiltonian given by:
\begin{align}
H &= \int d^2x\bigg[\frac{1}{2}g_b\sum_{n=1}^3\epsilon_{ik}\epsilon_{j\ell}b^i_nb^j_n\partial^k\phi\partial^\ell\phi - g_s\cos\bigg(\frac{2\pi}{6}\phi\bigg)\bigg]\nonumber\\
&=\int d^2x\bigg[\frac{1}{2}J(\partial_i\phi)^2 - g_s\cos\bigg(\frac{2\pi}{6}\phi\bigg)\bigg],
\end{align}
where $J\equiv\frac{3}{2} a^2 g_b$.  (For non-triangular lattices, we still have $J\sim a^2g_b$, but the numerical prefactor may change.)  This is a conventional sine-Gordon model (with some anisotropy for non-triangular lattices) which arises in the study of the conventional BKT transition.  The conventional gradient elasticity encodes logarithmic interaction between disclinations in a hexatic, screened by the proliferated dislocations down from the quadratic interaction found in a crystal, as we describe in detail in Appendix D.  This scalar sine-Gordon model thus describes the entropy-driven proliferation of disclinations with  increasing temperature, predicting the conventional BKT transition between  hexatic and isotropic fluids.  This completes the reproduction of the theory of two-stage classical melting of two-dimensional crystals, using the duality framework.

\begin{figure*}[t!]
 \centering
 \includegraphics[scale=0.23]{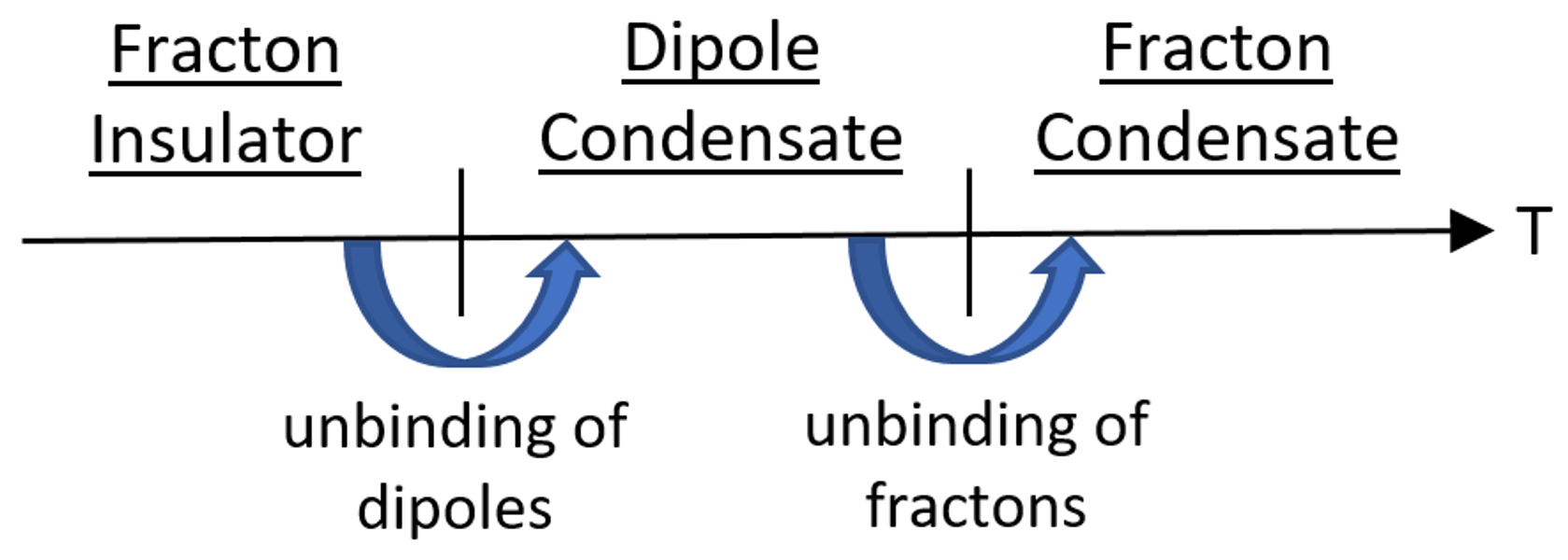}
 \caption{As the temperature is raised, a two-dimensional fracton tensor gauge theory exhibits a dipole unbinding transition, analogous to the solid-hexatic transition of elasticity theory.  At a higher temperature, the system will then undergo a fracton unbinding transition, analogous to the hexatic-liquid transition.}
 \label{fig:transitions}
 \end{figure*}

As another closely related application of the duality, we can immediately utilize the finite-temperature phase diagram of a two-dimensional crystal to predict thermal phases of two-dimensional fractons.  The duality predicts that, besides the fracton insulator, the tensor gauge theory should also admit two finite-temperature phases, distinguished by the proliferation of dipoles and fractons.  The fracton insulator should therefore undergo two phase transitions as the temperature is raised: unbinding of dipoles, followed by unbinding of fractons, as summarized in Fig.\ref{fig:transitions}.

\subsection{Ginzburg-Landau Theory of Tensor Superconductors}

As we have found in Sec. \ref{fgd}, the zero-temperature duality has predicted a description of a quantum crystal as a charged fractonic matter coupled to tensor gauge field electrodynamics. By construction we expect this novel field theory must provide a dual description of familiar quantum crystal and fluid phases. It is constructive to explore this novel gauge theory, a tensor Abelian-Higgs model “superconductor”, to clarify how it captures these conventional phases. In addition, one may hope that this dual description may provide access to new phases that are inaccessible (or not naturally described) in a direct description. Therefore, here we turn to the analysis of the tensor electrodynamics coupled to bosonic charged matter.  To this end, as with the conventional Abelian-Higgs model, it is convenient to soften the magnitude of the field constraints and generalize the dual model from Section \ref{fgd} to its equivalent Ginzburg-Landau formulation, coupled to a tensor gauge field theory.  We then use this simpler formulation to discuss the corresponding fracton phase diagram.  Here we limit our analysis to defining the model and outlining its general features, leaving a detailed study of the phases and phase transitions to the future.

We focus primarily on the study of dipole-condensed phases, where we expect to find the most novel physics.  Subsequent fracton condensation is expected to be equivalent to more conventional Ginzburg-Landau transitions, just as the hexatic-to-liquid transition can be understood in terms of a standard BKT analysis.  To describe a dipole-condensed phase, we introduce a complex order parameter $\psi_p$ for each minimal dipole species $p$, corresponding to the condensate strength of that species.  For example, on a square lattice, there will be two complex order parameters corresponding to the two minimal Burgers vectors, $\hat{x}$ and $\hat{y}$, while there are three order parameters required for a triangular lattice.

To construct the quantum Ginzburg-Landau theory of fractonic dipoles, we recall that the effective gauge field seen by a dipole $p^j$ takes the form $-p_j A^{ij}$, where $A^{ij}$ is a symmetric tensor gauge field.  (Note not to confuse the dipole $p_i$ with the momentum $k_i$.)  Restricting to the case of a square lattice, with two species of dipoles, the most general Lagrangian density allowed by symmetries, to fourth order in the two order parameters $\psi_1$ and $\psi_2$, takes the form:
\begin{align}
\mathcal{L} = i\sum_p \psi^\dagger_pD_0\psi_p - \frac{1}{2m}\sum_p|\Pi^{ik}_{\perp_p}D_i\psi_p|^2 \nonumber\\
 - \frac{\alpha}{2}\sum_p |\psi_p|^2 - \frac{\beta}{4}\sum_p|\psi_p|^4 - \frac{\beta'}{2}|\psi_1|^2|\psi_2|^2\nonumber\\
 +\frac{1}{2}\tilde{C}^{-1}_{ijk\ell}E_\sigma^{ij}E_\sigma^{k\ell} - \frac{\gamma}{2}B^iB_i,
\end{align}
where the magnetic field is a vector quantity, $B_i = \epsilon^{jk}\partial_jA_{ki}$, the covariant derivatives are defined as $D_0 = \partial_t + ip^i\partial_i\phi$ and $D_i = \partial_i + ip^jA_{ij}$, and the projector transverse to the dipole $p$ is $\Pi^{ik}_{\perp_p} = \delta^{ik} - \frac{p^ip^k}{p^2}$, as introduced in Reference \onlinecite{potter}.  This projection reflects the symmetry-enforced mobility restriction that dipoles can only move in the direction perpendicular to $p^i$.  The parameters $\alpha$, $\beta$, and $\beta'$ are real constants, where $\beta$, $\beta'>0$, while $\alpha$ is a tuning parameter which is negative in the condensed phase.  The constant $m$ represents the effective mass of dipoles in their allowed direction of motion.

By varying the action with respect to the two fields $\psi_1$ and $\psi_2$, we obtain the following equations of motion:
\begin{equation}
\bigg[iD_0 + \frac{1}{2m}\Pi_{jk}^{\perp_{p_1}} D_jD_k - \frac{\alpha}{2} - \frac{1}{2}(\beta|\psi_1|^2 + \beta'|\psi_2|^2)  \bigg]\psi_1 = 0,
\end{equation}
\begin{equation}
\bigg[iD_0 + \frac{1}{2m}\Pi_{jk}^{\perp_{p_2}} D_jD_k - \frac{\alpha}{2} - \frac{1}{2}(\beta|\psi_2|^2 + \beta'|\psi_1|^2)  \bigg]\psi_2 = 0.
\end{equation}
When $\alpha > 0$, the order parameters fluctuate around an energy minimum at $\psi_1=\psi_2=0$, indicating that no condensation has taken place.  When $\alpha<0$, however, at least one of the order parameters picks up a nonzero expectation value.  To determine the precise form of the condensation, we rewrite the potential portion of the Ginzburg-Landau action as:
\begin{equation}
V = \frac{\alpha}{2}|\Psi|^2 + \frac{\beta}{4}|\Psi|^4 +\frac{1}{2}(\beta'-\beta)|\psi_1|^2|\psi_2|^2
\label{glpot}
\end{equation}
where $|\Psi|^2 \equiv |\psi_1|^2 + |\psi_2|^2$.  If $\beta > \beta'$, then it is favorable for both order parameters to pick up the same nonzero expectation value, $|\langle\psi_1\rangle| = |\langle\psi_2\rangle| = \psi_0$, where:
\begin{equation}
\psi_0 = \sqrt{\frac{|\alpha|}{\beta+\beta'}},
\end{equation}
We will later briefly consider the case where $\beta < \beta'$.  For now, however, we proceed with the assumption $\beta > \beta'$.  In this case, we can further use the equations of motion to immediately read off the coherence length as:
\begin{equation}
\xi = \sqrt{\frac{1}{2m|\alpha|}},
\end{equation}
which represents the length scale on which the order parameters heal to their equilibrium value in the presence of a perturbation.

We can also vary the Lagrangian with respect to $\phi$, which yields the generalized Gauss's law of the theory as:
\begin{equation}
\partial^i\partial^j \tilde{C}^{-1}_{ijk\ell} E^{k\ell}_\sigma = -\sum_p p^i\partial_i(\psi^\dagger_p\psi_p),
\end{equation}
representing the contribution of a nonuniform distribution of dipoles to the charge density.  Similarly, we can vary with respect to $A_{ij}$, which yields the generalized Ampere's law:
\begin{align}
\partial_t \tilde{C}^{-1}_{ijk\ell}E^{k\ell}& + \gamma(\epsilon_{ik}\partial^kB_j + \epsilon_{jk}\partial^kB_i) \nonumber\\
&= -\frac{1}{2m}\sum_p (p_j\Pi_{ik}^{\perp_p} + p_i\Pi_{jk}^{\perp_p})\textrm{Im}(\psi_p^\dagger D^k\psi_p),
\end{align}
where the right-hand-side represents the tensor current carried by the dipoles.  If we now assume that we are in the dipole-condensed phase, such that $\psi_1 = \psi_0 e^{i\varphi_1}$ and $\psi_2 = \psi_0 e^{i\varphi_2}$, and keeping only lowest-order terms in fluctuations around the energy minimum, we can rewrite this equation of motion as:
\begin{align}
\partial_t \tilde{C}^{-1}_{ijk\ell}&E^{k\ell} + \gamma(\epsilon_{ik}\partial^kB_j + \epsilon_{jk}\partial^kB_i) \nonumber\\
&= -\frac{\psi_0^2}{2m}\sum_p (p_j\Pi_{ik}^{\perp_p} + p_i\Pi_{jk}^{\perp_p})(\partial_k\varphi_p + p_\ell A_{k\ell}).
\label{gleq}
\end{align}
From the structure of projectors, it is clear that only the off-diagonal $A_{12}$ component appears on the right-hand-side, indicating that only this component has picked up a mass.  This results in one of the two gapless modes of the insulating phase becoming gapped via the Higgs mechanism.  More specifically, plugging in the forms for $E_{ij}$ and $B_i$ in terms of $\phi$ and $A_{ij}$ into Eq. \ref{gleq}, picking the gauge where $\phi = 0$, and allowing $A_{12}$ to ``eat" the Goldstones modes ($\partial_1\varphi_2$ and $\partial_2\varphi_1$), the equations of motion for $A_{ij}$ can be written in Fourier space as:
\begin{align}
\begin{pmatrix}
c_1\omega^2 - q_y^2 & c_2\omega^2 & q_xq_y \\
c_1\omega^2 & c_2\omega^2 - q_x^2 & q_xq_y \\
-q_xq_y & -q_xq_y & -(4c_3\omega^2 - q^2 - \frac{2\psi_0^2}{m}p^2)
\end{pmatrix}
\begin{pmatrix}
A_{11} \\
A_{22} \\
A_{12}
\end{pmatrix}\nonumber\\ = 0,
\end{align}
where the three independent coefficients of the elastic tensor are written as $\tilde{C}^{-1}_{11,11} = \tilde{C}^{-1}_{22,22} \equiv c_1$, $\tilde{C}^{-1}_{11,22} = \tilde{C}^{-1}_{22,11}\equiv c_2$, and $\tilde{C}^{-1}_{12,12} = \tilde{C}^{-1}_{12,21} = \tilde{C}^{-1}_{21,12} = \tilde{C}^{-1}_{21,21} \equiv c_3$.  This equation has two independent solutions.  In the simplest limit, with $c_1 = c_2 = c_3\equiv c$, we obtain the following dispersion relations (to lowest order in $q^2$):
\begin{align}
\omega_{\pm}^2 = \frac{1}{16c}\bigg(2P + &q^2(3-\cos(4\theta))\nonumber\\
&\pm 2\sqrt{P(P+q^2 + q^2\cos(4\theta))}\bigg),
\end{align}
where we have used the polar representation $q_x = q\cos\theta$, $q_y = q\sin\theta$, and we have defined $P = 2\psi_0^2p^2/m$.  The mode with frequency $\omega_+$, corresponding to $A_{12}$, picks up a gap given by $\sqrt{P/4c}$.  The $A_{12}$ component alone will exhibit a Meissner effect, only being able to penetrate into the bulk of the system up to a scale given by:
\begin{equation}
\lambda = \sqrt{\frac{\gamma m}{2p^2|\psi_0|^2}} = \sqrt{\frac{\gamma m(\beta+\beta')}{2p^2\alpha}}.
\end{equation}
Meanwhile, the mode with frequency $\omega_-$, corresponding to the trace $A^i_i$, does not pick up a mass term and remains gapless.  This also means that there is no Meissner effect for $A^i_i$, which can penetrate into the interior of the system.

We can further use this framework to find the critical magnetic field which destroys the dipole superconducting phase.  We can do this by equating the energy difference between the superconducting and normal phases with the magnetic energy:
\begin{equation}
F_S - F_N = -\frac{\gamma}{2}B_c^2.
\end{equation}
We can then immediately write the critical field as:
\begin{equation}
B_c = \sqrt{\frac{\alpha^2}{\gamma(\beta+\beta')}}.
\end{equation}
There are many other interesting questions to be addressed regarding the dipole-condensed phase, such as the role and structure of vortex solutions.  We leave these important questions as topics of future research.

Before leaving the topic of Ginzburg-Landau theory, we return to the idea of another type of dipole condensation on the square lattice which does not respect the symmetries of the lattice.  Specifically, when $\beta < \beta'$, Equation \ref{glpot} indicates that it is energetically favorable to have only one of $\psi_1$ or $\psi_2$ pick up an expectation value, such that only one species of fundamental dipole has condensed.  In elasticity language, such a phase corresponds to a smectic, which breaks translational order in only one direction.  As in the case of a full dipole condensate, such a unidirectional dipole condensate leads to a mass for the $A_{12}$ component of the gauge field, gapping out one of the gauge modes, though the details of the dispersion will be different.  Various further details of the unidirectional dipole condensate are left to future investigations.

\subsection{Connection to Topological Crystalline Insulators}

Topological insulators\cite{kane,bernevig2,konig,fu,tireview,chong,chong2} (TIs) are a particular example of a broader class of systems known as symmetry protected topological (SPT) phases.\cite{senthil,xie}  These quantum phases of matter are best characterized in terms of their entanglement properties.  Specifically, an SPT phase cannot be disentangled to a direct product state without either breaking symmetry or undergoing a phase transition.  However, disentanglement becomes possible in the presence of symmetry-breaking perturbations.  The earliest work on TIs and other SPT phases focused on internal symmetries, such as time reversal or particle number conservation.  It was later realized that SPT phases could also be protected by crystalline symmetries.  Topological insulators protected by such spatial symmetries are known as topological crystalline insulators (TCIs).\cite{tci,tcireview}

As with other TIs, early studies of TCIs focused on the band theory of non-interacting electrons.  However, a robust classification and characterization of TCIs must account for interactions, a significantly more challenging problem.  Powerful tools have been developed for studying interacting SPT phases.  One, developed in the context of internal symmetries, is to consider gauging the symmetry protecting the SPT phase.\cite{gauging}  The resulting system will have long-range entanglement, described by a gauge theory with gauge group equivalent to the symmetry group of the original SPT phase.  (For a discrete internal symmetry group, we would say that the gauged system has intrinsic topological order.)  Furthermore, different SPT phases within the same Hilbert space ($e.g.$ a system of electrons subject to the same symmetries) map onto different long-range-entangled systems under gauging.  By studying the resulting gauge theories, one can thereby characterize and classify SPT phases, in a way which is robust to the introduction of interactions.

The theory of interacting SPT phases protected by internal symmetries, studied via gauging procedures and other techniques, is by now well-developed.  On the other hand, interacting SPT phases protected by crystal symmetries have only been studied relatively recently\cite{hiroki,hao,sj1,meng,sungjoon,zou,sj2}, and the set of available tools has been more limited.  Nevertheless, we expect that the gauging procedure applied to simpler SPT phases can be adapted to the case of crystal symmetries by the identification of a corresponding gauge ``flux."  A key insight which has recently been developed is that inserting a flux of a lattice symmetry is in an appropriate sense equivalent to inserting a topological lattice defect.\cite{else}  For example, in a two-dimensional crystal, a dislocation corresponds to a flux of translational symmetry, while a disclination corresponds to a flux of rotational symmetry.

Discussion on this topic so far has focused on promoting this symmetry flux to a non-dynamical gauge flux.  However, a fully gauged crystalline symmetry results in a dynamical gauge theory with dynamical lattice defects, $i.e.$ an elasticity theory.  As we have described in this work, such an elastic system can usefully be regarded as a fracton theory.  We therefore conclude that a gauged crystalline symmetry gives a state with fracton order.  By similar principles to the more conventional SPT physics, we expect that different TCIs should map onto different fracton phases under gauging.  By studying the resulting fracton phases, we should thereby be able to understand and classify the TCIs from which they are obtained.

In this way, the classification of topological crystalline insulators maps onto the problem of classification of fracton phases, which has numerous aspects.  First of all, different fracton phases have particles with different degrees of mobility, such as one-dimensional versus fully mobile dipoles.  And even for phases with the same particle mobility, there may be different quantum statistics associated with the various particle species.\cite{hanlayer}  As a further complication, in the event that there are additional symmetries in the problem besides the crystalline ones, one must have a complete classification of \emph{symmetry-enriched} fracton phases.  In such systems, one must ask several additional questions, such as how the symmetries act on the particle species.  It has also been shown that symmetry enrichment can cause extra mobility restrictions on a fracton theory, beyond those dictated by the gauge conservation laws.\cite{prl2,cheng,potter}  As such, a full understanding of symmetry-enriched fracton phases requires a systematic understanding of the ways in which global symmetries can restrict mobility.

The program outlined above leads to a direct mapping between two seemingly different physical problems, the classification of topological crystalline insulators and the classification of fracton phases (possibly with symmetry enrichment).  Methods of understanding both of these problems are still being actively developed, and it is therefore useful to have this dual perspective.  Advances in TCI physics may shed important light on the classification of fracton physics, and vice versa.  We leave the details of implementing this program as a task for the future.

\section{Experimental Signatures}

The analysis of this paper has shown how the well-established properties of elasticity theory can be equivalently reformulated in the useful new framework of fracton tensor gauge theories.  For example, we have used this dual language to reproduce the mobility restrictions of topological lattice defects in terms of a simple set of higher moment conservation laws.  But in addition to the reproduction of known facts, it is important to establish whether or not our duality makes any new predictions for crystalline phases or opens up new sort of questions which were not apparent in the conventional formulation of elasticity theory.  In this section, we discuss several clear indications of fracton physics in crystals which, to the best of our knowledge, have not been studied within the conventional framework of elasticity theory.  Undoubtedly many other fascinating experimental implications remain to be explored.

\subsection{Pinch Point Singularities}

\begin{figure}[t!]
 \centering
 \includegraphics[scale=0.42]{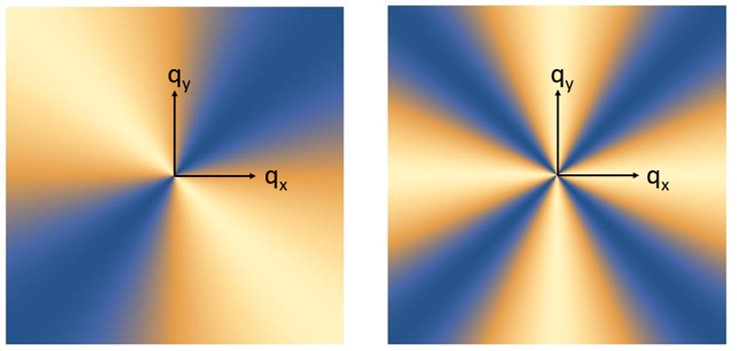}
 \caption{At left is a schematic plot of $\langle E^x(q)E^y(-q)\rangle$ in the $q_x$-$q_y$ plane, displaying the characteristic two-fold pinch-point singularity of conventional gauge theories.  At right is an analogous plot of $\langle E^{xx}(q)E^{yy}(-q)\rangle$, displaying the four-fold pinch-point singularity of a rank-2 tensor gauge theory.  Figure adapted from Reference \onlinecite{pinch}.}
 \label{fig:pinch}
 \end{figure}

One key signature of the presence of an emergent $U(1)$ gauge field is the existence of certain characteristic singularities in certain physical correlation functions.  For example, in Coulomb spin ice materials, it has been shown that the spin-spin correlation functions map directly onto electric (or magnetic) field correlators of an emergent Maxwell gauge field:
\begin{equation}
\langle S^z(q) S^z(-q)\rangle = \sum_{ij} C_{ij} \langle E^i(q)E^j(-q)\rangle,
\end{equation}
for some coefficients $C_{ij}$ dictated by the symmetry of the lattice.  Importantly, the Maxwell electric field correlator has characteristic singular behavior:
\begin{equation}
\langle E^i(q)E^j(-q)\rangle\sim \delta^{ij} - \frac{q^iq^j}{q^2}.
\end{equation}
The components of this correlator are singular in the sense that the $q\rightarrow 0$ limit depends upon the direction in which $q=0$ is approached, as depicted in Figure \ref{fig:pinch}.  This sort of ``pinch point" singularity is readily observed in spin-spin correlation functions, which can be measured via neutron scattering experiments.\cite{henley1,henley2}  Indeed, such singularities have been observed in certain spin ice materials, serving as a clear indication of the presence of an emergent gauge theory.\cite{fennell}

Similar sorts of singularities are expected for theories featuring emergent fracton tensor gauge theories, as discussed recently in the context of spin models hosting fracton excitations.\cite{pinch}  It was shown that tensor gauge theories give rise to pinch-point singularities with a characteristic four-fold symmetry, in contrast with the two-fold symmetry of pinch-point singularities in more conventional gauge theories, as shown in Figure \ref{fig:pinch}.  In the context of elasticity theory, these singularities can be found in the stress-stress correlation functions.  Relying on the results of Reference \onlinecite{pinch}, we can immediately write the low-energy stress-stress correlator as:
\begin{align}
\langle\sigma^{ij}(q)\sigma^{k\ell}(-q)\rangle &= \epsilon^{ia}\epsilon^{jb}\epsilon^{kc}\epsilon^{\ell d}\langle E_{ab}(q)E_{cd}(-q)\rangle\nonumber\\
&\sim \frac{1}{2}(\delta^{ik}\delta^{j\ell} + \delta^{i\ell}\delta^{jk}) \nonumber\\
&\,\,\,\,\,\,\,\,\,- \epsilon^{ia}\epsilon^{jb}\epsilon^{kc}\epsilon^{\ell d} \frac{q_aq_bq_cq_d}{q^4}.
\end{align}
Not all components of this correlator have singular behavior.  But if we measure, for example, $\langle \sigma^{xx}(q)\sigma^{yy}(-q)\rangle$, it will exhibit precisely the sort of singularity depicted in Figure \ref{fig:pinch}.  The same sort of singularities will also manifest directly in correlation functions of the lattice displacement, in the form of the strain-strain correlator:
\begin{equation}
\langle u_{ij}(q)u_{k\ell}(-q)\rangle = C^{-1}_{ijnm}C^{-1}_{k\ell rs}\langle\sigma^{nm}(q)\sigma^{rs}(-q)\rangle.
\end{equation}
This correlator will also feature the four-fold singularities of Figure \ref{fig:pinch}, though which components are singular will depend on the symmetry of the underlying lattice.  This experimentally accessible quantity provides a clear signature of an emergent tensor gauge structure in the theory of elasticity, making a precise connection with the physics of fractons.

\subsection{Absence of Zero Modes}

Another important signature of fracton physics arises when we consider a crystal of disclinations on top of the background lattice.  Such a disclination crystal is encountered, for example, when a crystalline medium is wrapped into a spherical shape, such that the associated topology necessitates the existence of disclinations.  Specifically, a hexagonal lattice wrapped into a sphere is required to have a crystal of at least twelve 5-fold disclinations.  Let us choose the density of disclinations such that the disclination crystal is incommensurate with the underlying crystal.  In this case, the original crystal and disclination crystal represent distinct forms of symmetry breaking, which naively would involve two separate sets of gapless phonon modes.  For two regular coexisting incommensurate crystalline structures, the joint system would indeed have two independent sets of gapless phonons.  However, for a system of disclinations, which behave as fractons, there are no matrix elements for motion of the disclination crystal.  Such processes are ruled out by the conservation laws of the theory, which prevents gapless phonons from arising as ``soft" modes translating the disclination crystal.  We therefore conclude that a disclination crystal will not have all of the gapless Goldstone modes which would naively be required based on broken symmetries.  This could be verified, for example, by measuring correlations of the displacement $u_d$ of disclinations relative to their underlying crystal.  Correlations such as $\langle u_d(x)u_d(0)\rangle$ should have exponentially short-ranged behavior, as opposed to the long-ranged correlations which would be expected in the presence of Goldstone modes.

\subsection{Disclination Mobility in Finite-Temperature Hexatics}

While our dual formulation of elasticity theory predicts that disclinations are immobile in a crystal, this restriction is lifted in a hexatic phase.  While a fracton still cannot move by itself, it can move through absorption of a dipole ($i.e.$ a dislocation).  In a hexatic phase, dislocation defects have proliferated throughout the system, and a disclination can move through interaction with this finite density of dislocations.  For simplicity, let us consider a finite-temperature hexatic phase, driven by the conventional Halperin-Nelson thermal unbinding of dislocations.  In such a phase, a disclination will effectively undergo a random walk through the frequent absorption of randomly directed dislocations, causing the disclination to diffuse through the system.  The rate of this diffusion is directly set by the density of dislocations.  The typical velocity of a disclination is directly proportional to the dislocation density, $n_d$.\cite{screening}  In turn, this leads to an effective fracton diffusion constant which is also directly proportional to dislocation density:
\begin{equation}
D_f\sim n_d(T),
\end{equation}
From our earlier analysis, we can conclude that a thermal hexatic phase has a dislocation density given by $n_d(T) \sim \xi_1^{-2}(T)$, where $\xi_1(T)$, as defined in Eq. \ref{thermlength}, has strong $T$ dependence near the transition, but asymptotes to a constant at higher temperatures.  This provides a clear prediction for the temperature dependence of disclination mobility in the hexatic phase, which in principle can be directly detected in experiments.

\section{Conclusions}

In this work, we have demonstrated a duality between elasticity of a two-dimensional crystal and a $U(1)$ fracton tensor gauge theory, in a natural tensor analogue of conventional particle-vortex duality.  The topological lattice defects of elasticity theory map onto the charges of the gauge theory, with disclinations as fractons and dislocations as dipoles, while the two phonon modes map onto the gapless gauge modes of the gauge theory, as summarized in Figure \ref{fig:dictionary}.  This duality provides numerous insights into $U(1)$ fracton physics based on well-established results of elasticity theory.  Further connections with three-dimensional $Z_n$ fracton lattice models may then be possible via the Higgs mechanism.\cite{higgs1,higgs2}  For example, our physical picture of phase transitions in fracton systems, in terms of unbinding of dipoles and fractons, may shed light on quantum phase transitions in the gapped fracton models.  In turn, the fracton tensor gauge theory allows for a convenient reformulation of several aspects of elasticity theory, such as the restricted mobility of lattice defects.  Our work has numerous other implications, such as drawing a connection between fractons and interacting topological crystalline insulators.  This duality opens the door for the future exchange of ideas between the new field of fractons and more-established ideas in the field of elasticity.

\section*{Acknowledgments}

The authors acknowledge useful conversations with Yang-Zhi Chou, Han Ma, Matthew Fisher, Shriya Pai, Abhinav Prem, Albert Schmitz, Rahul Nandkishore, Senthil Todadri, Drew Potter, Ajesh Kumar, Brian Swingle, Yizhi You, and Mike Hermele.  We also thank Duncan Haldane for discussion and for emphasizing to us the difficulty of imposing tracelessness of $E$ and $E_\sigma$ in a continuum tensor gauge theory.  This work was supported by the Simons Investigator Award to LR from the Simons Foundation and partially by the NSF Grant 1734006.  LR also thanks the KITP for its hospitality as part of the Fall 2016 Synthetic Matter workshop and sabbatical program, during which this work was initiated and supported by the NSF grant PHY-1125915.

\section*{Appendix A: Particle-Vortex Duality}

In the main text, we have described a tensor version of particle-vortex duality.  For readers unfamiliar with conventional particle-vortex duality (sometimes called ``Dasgupta-Halperin duality"), we here review some of its basic aspects, which describes a two-dimensional superfluid in terms of a Maxwell $U(1)$ gauge theory coupled to charged matter.  The major hint for this duality comes from examining the excitation spectra of both theories.  A two-dimensional superfluid features a gapless Goldstone mode plus logarithmically interacting vortices.  Similarly, the two-dimensional $U(1)$ gauge theory features a gapless photon coupled to logarithmically interacting charges.  It therefore seems reasonable that an appropriate duality transformation will map the Goldstone mode onto the photon, while vortices map onto charges.

To see this duality explicitly, it is easiest to start on the gauge theory side, which simply consists of a $U(1)$ gauge field $a_i$ coupled to charged particles.  The gauge field itself is governed by a conventional Maxwell Hamiltonian:
\begin{equation}
H = \int d^2x\,\frac{1}{2}(e^ie_i + b^2),
\end{equation}
where $b = \epsilon^{ij}\partial_ia_j$.  The charges are gapped, and the precise form of their action is unimportant.  The only important piece of physics from the charge sector is Gauss's law, relating charge density to the electric field $e^i$ conjugate to $a^i$:
\begin{equation}
\partial_ie^i = 2\pi\rho,
\end{equation}
where the normalization has been chosen for later convenience.

Within this charge-free sector, the electric field obeys the source-free condition, $\partial_ie^i = 0$.  We can therefore conveniently describe the gapless photons by rigidly enforcing this constraint, which has the general solution:
\begin{equation}
e^i = \epsilon^{ji}\partial_j\phi,
\label{super}
\end{equation}
for scalar field $\phi$.  Since $a_i$ generates translations of $e_i$, it is easy to check that $b$ generates translations of $\phi$.  In other words, $b$ is the canonical conjugate to $\phi$, which we relabel as $n = b$.  In terms of these variables, the low-energy Hamiltonian becomes:
\begin{equation}
H = \int d^2x\,\frac{1}{2}((\partial_i\phi)^2 + n^2),
\end{equation}
which is precisely the effective theory for the Goldstone mode of a superfluid, where $\phi$ is interpreted as the phase angle of the superfluid condensate, and $n$ corresponds to the boson number variable.  This indicates that the photon of two-dimensional Maxwell theory can be mapped directly onto the superfluid Goldstone mode.

In order to complete the particle-vortex duality, we must also match the gauge theory's charges (``particles") with the vortices of the superfluid.  To see this, we can write the total charge enclosed in a region of space $V$ (with boundary $\partial V$) as:
\begin{align}
q = \int_V d^2x\,\rho = \frac{1}{2\pi}\int_V d^2x\,\partial_ie^i = \frac{1}{2\pi}\int_{\partial V} dn_i e^i.
\end{align}
Plugging in the low energy form for $e^i$ from Equation \ref{super}, we obtain:
\begin{equation}
q = \frac{1}{2\pi}\oint_{\partial V} d\ell^i\partial_i\phi = \frac{\Delta\phi}{2\pi},
\end{equation}
so a unit charge represents a $2\pi$ winding of $\phi$ around the curve.  In other words, a charge of the gauge theory corresponds to a vortex of the superfluid.

With this correspondence in place, we have now matched the complete spectra between the superfluid and the $U(1)$ gauge theory, with vortices acting as charges and the Goldstone mode playing the role of the photon.  The gauge theory formulation provides a convenient dual description of the superfluid.  Furthermore, the duality remains valid even as the superfluid transitions to a gapped Mott insulating phase.  An insulator of bosons can equivalently be thought of as a condensate of vortices.  On the gauge theory side, this corresponds to the condensation of charges, which gaps the photon via the Anderson-Higgs mechanism.  The $U(1)$ gauge theory is thereby able to capture the full phase diagram of a two-dimensional boson system.

\section*{Appendix B: Duality from Fractons to Elasticity}

We have already shown how to map elasticity theory directly onto a fracton tensor gauge theory.  To obtain the duality in the reverse direction, one could in principle simply reverse each step of the previous derivation.  However, this process is cumbersome and does not allow for immediate generalization to other tensor gauge theories.  Luckily, there is a shortcut which allows any gauge theory to be quickly converted to its dual formulation.  Within the gapless gauge sector, where there are no charges, the system obeys the source-free Gauss's law:
\begin{equation}
\partial_i\partial_jE^{ij} = 0,
\label{sourcefree}
\end{equation}
and the Hamiltonian of the system takes the form:
\begin{equation}
H = \int d^2x\,\frac{1}{2}(\tilde{C}^{ijk\ell}E_{ij}E_{k\ell} + B^iB_i).
\end{equation}
Within this sector, any configuration of $E_{ij}$ can be written in terms of the general solution to the source-free Gauss's law, which takes the form:
\begin{equation}
E_{ij} = \frac{1}{2}(\epsilon_{ik}\epsilon_{j\ell}\partial^k u^\ell + \epsilon_{jk}\epsilon_{i\ell}\partial^k u^\ell) = \epsilon_{ik}\epsilon_{j\ell}u^{k\ell},
\label{sol}
\end{equation}
for arbitrary vector $u^i$ (and corresponding strain tensor $u^{ij}$).  It is easy to check that, since $A_{ij}$ is a field canonically conjugate to $-E_{ij}$, the canonical conjugate to $u_i$ is:
\begin{equation}
\pi_i = \epsilon_{ij}B^j.
\end{equation}
In terms of these new dual variables, the Hamiltonian takes the form:
\begin{align}
&H = \int d^2x\,\frac{1}{2}(C^{ijk\ell}u_{ij}u_{k\ell} + \pi^i\pi_i),
\end{align}
which is precisely the Hamiltonian for two-dimensional elasticity theory.

Now that we have established a duality within the gapless sector, we must also convert the charges of the gauge theory into elasticity language.  In any region of space $V$ with boundary $\partial V$, we can write the total enclosed charge as:
\begin{align}
q &= \int_V d^2x\,\rho = \int_V d^2x\,\partial_i\partial_j E^{ij} \nonumber\\
&= \int_{\partial V} dn_i\,\partial_j E^{ij} = -\oint_{\partial V} d\ell^i \epsilon_{ik}\partial_j E^{kj}.
\end{align}
On the boundary, away from any charges, we can then plug in the low-energy form for $E^{ij}$ from Equation \ref{sol}, yielding:
\begin{equation}
q = \frac{1}{2}\oint_{\partial V} d\ell^i \partial_i(\epsilon_{jk} \partial^j u^k) = \frac{1}{2}\Delta(\epsilon_{jk}\partial^j u^k).
\end{equation}
A charge on the gauge theory side therefore represents a winding of the bond angle $\theta_b = \epsilon_{jk}\partial^ju^k$ around some point, which is precisely the definition of a disclination defect.  The size of the fundamental charge in the gauge theory will be set by the minimal winding of $\theta_b$ around a curve, which depends on the symmetry of the lattice.  Specifically, in a $C_n$ symmetric lattice, the minimal winding is $\Delta(\epsilon_{jk}\partial^ju^k) = 2\pi/n$, such that the fundamental charge of the gauge theory is $q = \pi/n$.

There is one other important type of excitation which is present on both sides of the duality: dipoles of equal and opposite charges/disclinations.  On the elasticity side, this should correspond to a dislocation defect.  We can obtain the correspondence explicitly by considering the total dipole moment in some region $V$:
\begin{equation}
P^i = \int_V d^2x\, x^i\partial_j\partial_k E^{jk} = \int_{\partial V}dn_j\,(x^i\partial_k E^{jk} - E^{ij}).
\end{equation}
Plugging in the low-energy form of $E^{ij}$ from Equation \ref{sol} and rearranging a few terms, we obtain:
\begin{equation}
P^i = \oint_{\partial V} d\ell^j \partial_j\bigg(\epsilon^{ik}u_k - \frac{1}{2}x^i\epsilon^{k\ell}\partial_k u_\ell \bigg).
\end{equation}
Assuming that there are zero net charges (disclinations) contained inside the region, so that $\epsilon^{k\ell}\partial_k u_\ell$ does not wind around the closed curve, we will be left with:
\begin{equation}
P^i = \oint_{\partial V} d\ell^j\partial_j(\epsilon^{ik}u_k) = \epsilon^{ik}\Delta u_k = \epsilon^{ik}b_k,
\label{dipole}
\end{equation}
where $b_k$ is the Burgers vector.  We see that a dipole in the gauge theoretic language corresponds to a dislocation in elasticity language, with Burgers vector perpendicular to the dipole moment, as expected.  Finally, we note that the above relationship leads to a convenient alternative formulation of the duality which is useful for describing a system of dislocations, as discussed in Appendix C.

\section*{Appendix C: Duality without Disclinations}

As discussed in the main text, disclination defects are extremely energetically costly in the solid phase, and thus do not play an important role in the low-energy elastic theory of a crystal (though they play an important role in the hexatic phase and its transition to the isotropic liquid).  It is therefore reasonable to construct an effective low-energy theory for a crystal, with a corresponding tensor gauge dual, in which dislocations are the fundamental charges, without making any reference to disclinations at all.  In this case, we expect the charges of the gauge theory to all have a vector character, as opposed to the scalar charges considered earlier.  Thus, the charge sector of the theory should have two components.  The gapless phonon sector should also still have two components, giving four total local degrees of freedom.

Using these clues, we formulate a theory in terms a generic tensor $\tilde{A}_{ij}$, without any index symmetry, which has four independent components.  We call its canonical conjugate variable $\tilde{E}_{ij}$.  (We use tildes to distinguish from the symmetric tensors used in the main text.)  We stipulate that the Gauss's law constraint on this tensor is:
\begin{equation}
\partial_i \tilde{E}^{ij} = \rho^j,
\end{equation}
for vector charge density $\rho^j$.  Within the charge-free sector, the most general low-energy Hamiltonian for the gauge modes is:
\begin{equation}
H = \int d^2x\,\frac{1}{2}(\tilde{C}^{ijk\ell}\tilde{E}_{ij}\tilde{E}_{k\ell} + \tilde{B}^i\tilde{B}_i),
\end{equation}
where the magnetic field is given by
\begin{equation}
\tilde{B}^i = \epsilon_{jk}\partial^j \tilde{A}^{ki}.
\end{equation}
The source-free Gauss's law, $\partial_i\tilde{E}^{ij} = 0$, has the generic solution:
\begin{equation}
\tilde{E}^{ij} = \epsilon^{ik}\epsilon^{j\ell}\partial_k u_\ell.
\end{equation}
It is easy to check that $B_i$ is the canonical conjugate to $u_i$, which we label $\pi_i$.  In terms of these new variables, the Hamiltonian becomes:
\begin{equation}
H = \int d^2x\,(C^{ijk\ell}u_{ij}u_{k\ell} + \pi^i\pi_i),
\end{equation}
which once again takes the standard elastic form, describing two phonon modes.

We must also determine the correspondence between the vector charges and dislocations.  The total charge in region $V$ with boundary $\partial V$ is given by:
\begin{equation}
q^j = \int_V d^2x\,\rho^j = \int_V d^2x\,\partial_i\tilde{E}^{ij} = \int_{\partial V} dn_i\tilde{E}^{ij}.
\end{equation}
Plugging in the low-energy form for $\tilde{E}^{ij}$, we obtain:
\begin{equation}
q^j = \oint_{\partial V} d\ell^i\partial_i(\epsilon^{\ell j}u_\ell) = \Delta (\epsilon^{\ell j}u_\ell) = \epsilon^{\ell j}b_\ell.
\end{equation}
We therefore see that a vector charge indeed corresponds to a dislocation, as expected.  We have now matched the excitation spectrum on both sides of the duality: dislocations with vector charges and phonons with gauge modes.  This completes our duality of the phonon-dislocation theory.  As we have discussed, this theory does not incorporate disclinations.  But for low-energy purposes within the solid phase, the disclination-free treatment should be accurate.

\section*{Appendix D: Disclination Screening in the Hexatic Phase}

In the main text, we described how two-dimensional crystals and their corresponding fracton tensor gauge theories undergo two finite-temperature phase transitions, corresponding to the proliferation of dislocations (dipoles), followed by the proliferation of disclinations (fractons).  The first such transition is fairly simple to see in gauge theory language.  As we derived earlier in Equation \ref{limit}, the long-distance interaction potential between two dipoles takes a logarithmic form.  For two equal and opposite dipoles, $p$ and $-p$, this interaction take the form:
\begin{equation}
V(r)\sim \frac{p^2}{4\pi}\log r.
\end{equation}
This is the same type of interaction that occurs in a two-dimensional Coulomb gas, or between vortices in a superfluid.  As in those more familiar systems, a simple argument based on the free energy per particle indicates that the system will undergo a finite-temperature unbinding transition.  The energy of an isolated dipole grows as $p^2\log L$, while the entropy behaves as $T\log L$, yielding the free energy per dipole as:
\begin{equation}
F =\mathcal{E}-TS = (p^2-T)\log L.
\end{equation}
At zero temperature, the energy term dominates and $F>0$, so forming isolated dipoles is unfavorable.  At higher temperatures however, the free energy per particle becomes negative, $F<0$.  This will result in the proliferation of dipoles, resulting in a gauge theory analogue of the hexatic phase.

While the unbinding of dipoles is fairly easy to understand in the gauge theory language, fracton unbinding is slightly more subtle.  From our earlier potential formulation, we found that the energy of an isolated fracton in a solid grows as $L^2$, which would keep the fractons bound at all temperatures.  If fractons are to proliferate, there must be a mechanism which drastically reduces their energy within the gauge hexatic phase.  Precisely such a reduction occurs due to screening by the finite density of dipoles within this phase.  To see this, we write a self-consistent equation for the total electrostatic potential of a fracton, summing the contributions from the bare fracton and its screening cloud of dipoles, following the treatment of Reference \onlinecite{screening}:
\begin{equation}
\phi(r) = \phi_q(r) + \sum_p\int d^2r'\,n_p(T,\phi(r'))\phi_p(r-r').
\label{potsum}
\end{equation}
Here, $\phi_q$ is the bare fracton potential, $\phi_p$ is the potential generated by a dipole, and the sum runs over the fundamental dipole moments.  The density $n_p(T,\phi)$ represents the density of $p$-directed dipoles at temperature $T$ and potential $\phi$.  In the presence of a potential, the Boltzmann weights of dipoles shift, giving:
\begin{equation}
n_p = n_0e^{-\beta p^i\partial_i\phi}\sim n_0(1-\beta p^i\partial_i\phi),
\end{equation}
where $\beta = 1/T$ is the inverse temperature, $n_0$ is a finite background dipole density, and we have approximated that the perturbing potential is weak.  (This approximation breaks down close to the fracton, but captures the correct long-distance physics.)  Plugging this form into Equation \ref{potsum}, we obtain:
\begin{align}
\begin{split}
&\phi(r) = \phi_q(r) - \\
\sum_p& n_0\int d^2r'(1-\beta p^i\partial_i'\phi(r'))\frac{(p\cdot  (r-r'))}{4\pi}\log(r-r').
\end{split}
\end{align}
We now use the facts that $\sum_p p_i = 0$ and $\sum_p p_ip_j = \alpha\delta_{ij}$, where the value of $\alpha$ depends on the lattice under consideration, to rewrite the above equation as:
\begin{equation}
\phi(r) = \phi_q(r) + \frac{\alpha\beta n_0}{4\pi}\int d^2r'\partial_i'\phi(r')(r-r')^i\log(r-r').
\end{equation}
Taking a Fourier transform and solving for $\phi$, we obtain:
\begin{equation}
\phi(k) = \frac{k^2\phi_q(k)}{k^2 + \alpha\beta n_0} = \frac{q}{k^2(k^2 + \alpha\beta n_0)}.
\end{equation}
At small $k$, we have:
\begin{equation}
\phi(k)\sim \frac{q}{\alpha\beta n_0k^2}.
\end{equation}
This indicates that the long-distance behavior of the screened potential is:
\begin{equation}
\phi_{scr}(r)\sim \frac{q}{\alpha\beta n_0}\log r.
\end{equation}
We now see that, after accounting for screening by the proliferated dipoles of the hexatic phase, the energy of an isolated fracton will behave as $\log L$, instead of $L^2$.  With this reduction of energy, entropic effects will take over at a finite temperature, leading to an unbinding transition of fractons.  We thereby reach a finite-temperature phase in which both dipoles and fractons have proliferated.


\begin{thebibliography}{9}
\bibitem{foot5}
This tensor will also have a contribution from processes in which a disclination (fracton) moves while emitting or absorbing a dislocation (dipole).
\end{thebibliography}
\end{document}